\documentclass[12pt]{article}
\usepackage{epsf}
\usepackage{amsfonts}
\usepackage{amsmath}
\usepackage{lscape}
\usepackage{hyperref}
\usepackage{longtable}

\newcommand{\bmat}{\left(\begin{array}}
\newcommand{\emat}{\end{array}\right)}

\def\yzero{\smash{\hbox{$y\kern-4pt\raise1pt\hbox{${}^\circ$}$}}}

\def\beq{\begin{equation}}
\def\eeq{\end{equation}}
\def\beqa{\begin{eqnarray}}
\def\eeqa{\end{eqnarray}}

\def\-{\hphantom{-}}

\def\s2{\frac{1}{\sqrt2}}

\def\beq{\begin{equation}}
\def\eeq{\end{equation}}
\def\beqa{\begin{eqnarray}}
\def\eeqa{\end{eqnarray}}

\def\IF{\relax{\rm I\kern-.18em F}}
\def\II{\relax{\rm I\kern-.18em I}}
\def\IP{\relax{\rm I\kern-.18em P}}
\def\IC{\relax\hbox{\kern.25em$\inbar\kern-.3em{\rm C}$}}
\def\IR{\relax{\rm I\kern-.18em R}}

\def\Dsl{\,\raise.15ex\hbox{/}\mkern-13.5mu D} 
\def\IZ{Z\kern-.4em  Z}

\def\Ad{\bf \rm Ad}

\def\cH{{\cal H}}
\def\cS{{\cal S}}
\def\cC{{\cal C}}

\def\hri#1#2{\href{http://arxiv.org/abs/#1}{[ArXiv:#1]#2}}
\def\hre#1#2{\href{http://arxiv.org/abs/#1/#2}{[ArXiv:#1/#2]}}

\catcode`\@=11   
\newdimen\@rotdimen
\newbox\@rotbox  

\def\@vspec#1{\special{ps:#1}}
\def\@rotstart#1{\@vspec{gsave currentpoint currentpoint translate
   #1 neg exch neg exch translate}}
\def\@rotfinish{\@vspec{currentpoint grestore moveto}}
\def\@rotr#1{\@rotdimen=\ht#1\advance\@rotdimen by\dp#1%
   \hbox to\@rotdimen{\hskip\ht#1\vbox to\wd#1{\@rotstart{90 rotate}%
   \box#1\vss}\hss}\@rotfinish}
\def\@rotl#1{\@rotdimen=\ht#1\advance\@rotdimen by\dp#1%
   \hbox to\@rotdimen{\vbox to\wd#1{\vskip\wd#1\@rotstart{270 rotate}%
   \box#1\vss}\hss}\@rotfinish}%
\def\@rotu#1{\@rotdimen=\ht#1\advance\@rotdimen by\dp#1%
   \hbox to\wd#1{\hskip\wd#1\vbox to\@rotdimen{\vskip\@rotdimen
   \@rotstart{-1 dup scale}\box#1\vss}\hss}\@rotfinish}%
\def\@rotf#1{\hbox to\wd#1{\hskip\wd#1\@rotstart{-1 1 scale}%
   \box#1\hss}\@rotfinish}%
\def\rotate{\@ifnextchar[{\@rotate}{\@rotate[l]}}
\def\@rotate[#1]#2{\setbox\@rotbox=\hbox{#2}\@nameuse{@rot#1}\@rotbox}
\def\hre#1#2{\href{http://arxiv.org/abs/#1/#2}{[ArXiv:#1/#2]}}

\catcode`\@=12

\topmargin
-1.5cm
\textwidth
15.5cm
\textheight
23.5cm
\oddsidemargin
0.7cm
\evensidemargin
1.2cm

\begin{document}

\makeatletter
\@addtoreset{equation}{section}
\makeatother
\renewcommand{\theequation}{\thesection.\arabic{equation}}
\pagestyle{empty}
\rightline{ IFT-UAM/CSIC-12-46}
\rightline{ Nikhef/2012-011}
\vspace{0.1cm}
\begin{center}
\LARGE{Discrete Gauge Symmetries in Discrete MSSM-like Orientifolds
 \\[10mm]}
\large{ L.E. Ib\'a\~nez$^{1,3}$ A.N. Schellekens$^2$
 and A. M. Uranga$^{3}$ \\[3mm]}
\footnotesize{
${}^{1}$ Departamento de F\'{\i}sica Te\'orica, M\'odulo 8, Facultad de Ciencias,\\[-0.3em]
Universidad Aut\'onoma de Madrid,
Cantoblanco, 28049 Madrid, Spain \\[2mm] 
 $^2$ Nikhef, Science Park 105, 1098 XG Amsterdam,
The Netherlands;\\
IMAPP, Radboud Universiteit, Nijmegen, The Netherlands;\\
Instituto de Fisica Fundamental (IFF-CSIC),\\
Serrano 123, Madrid 28006, Spain.\\[2mm]
${}^{3}$ Instituto de F\'{\i}sica Te\'orica UAM/CSIC, \\ 
c/ Nicol\'as Cabrera 13-15, 28049 Madrid, Spain\\
[4mm]}
\small{\bf Abstract} \\[5mm]
\end{center}
\begin{center}
\begin{minipage}[h]{17.0cm}
{\small
Motivated by the necessity of discrete ${\mathbb{Z}_N}$ symmetries in the MSSM to insure baryon stability, 
we study the origin of discrete gauge symmetries from open string sector $U(1)$'s in orientifolds based on rational conformal field theory.
By means of an explicit construction, we find an integral basis for the couplings of axions and
$U(1)$ factors for all simple current MIPFs and orientifolds of all 168 Gepner models, a total of
32990 distinct cases. We discuss how the presence of discrete symmetries surviving as a subgroup
of broken $U(1)$'s can be derived using this basis. We apply this procedure to models with MSSM chiral spectrum, concretely 
to all known $U(3)\times U(2) \times U(1)\times U(1)$ and $U(3)\times Sp(2) \times U(1)\times U(1)$ configurations
with chiral bi-fundamentals, but no chiral tensors, as well as some $SU(5)$ GUT models. 
We find examples of models with  
$\mathbb{Z}_2$ (R-parity) and $\mathbb{Z}_3$ symmetries that forbid certain $B$ and/or $L$ violating MSSM couplings.
Their presence is however relatively rare, at the level of a few percent of all cases.
}

\end{minipage}
\end{center}
\newpage
\setcounter{page}{1}
\pagestyle{plain}
\renewcommand{\thefootnote}{\arabic{footnote}}
\setcounter{footnote}{0}

\hyphenation{veri-fi-able}

\section{Introduction}

Discrete symmetries are often used in model building in order to argue that certain
otherwise allowed terms in the effective Lagrangian are in fact absent. 
So far in nature no such symmetry has been observed, apart from CPT. Furthermore, all 
allowed standard model interactions are indeed observed, with the exception of the QCD strong CP violating term, whose apparent
absence can however not be explained in terms of an exact discrete symmetry. 

On the other hand, it has been quite common to invoke discrete symmetries in model building beyond the standard model,
and especially supersymmetric model building. Indeed, a generic point in the full parameter space of the
supersymmetrized standard model with soft supersymmetry breaking terms would be in disagreement with
observation, most notably because of the allowed baryon and/or lepton number violating
$UDD$, $QLD$, $LLE$ and $LH_u$ terms\rlap,\footnote{Here $Q, U, D, L$ and $E$ denote the usual MSSM superfields, with $U, D$ and $E$ referring to anti-particles.  
The two Higgs superfields are denoted 
$H_u$ and $H_d$, and couple to up and down quarks respectively.} leading to dimension four operators. These give rise to catastrophic proton decay
rates if all  of them are present with coefficients of order one, and serious constraints even if some of them are absent.
The most common solution to this problem is to postulate a discrete
symmetry that forbids them, such as R-parity \cite{Farrar:1978xj} or Baryon-triality \cite{Ibanez:1991pr}.

Even if the dimension four operators are absent, one has to
worry\footnote{These operators would be less problematic if proton decay rates are suppressed non-perturbatively, as suggested recently in \cite{Martin:2011nd}. However, lattice
computations do not show such a suppression, see {\it e.g.} \cite{Aoki:2008ku} and references therein.} about 
dimension five operators originating from the superpotential terms $QQQL$ and $UUDE$, which preserve $B\!-\!L$ and
are therefore not forbidden by R-parity.  This is because R-parity may be considered as a ${\mathbb{Z}_2}$ subgroup of $B-L$.


In string theory one cannot simply postulate a discrete symmetry. It must be a verifiable property of a given string
realization. There are strong arguments supporting that global symmetries (either continuous or discrete) cannot exist in theories containing quantum gravity, rather they must be gauge symmetries (see e.g. \cite{Banks:2010zn} for a recent discussion, and \cite{Banks:1988yz,Abbott:1989jw,Coleman:1989zu,Kallosh:1995hi,Alford:1988sj,Krauss:1988zc,Preskill:1990bm} \ for earlier references). Hence, any exact discrete symmetry in string theory must be gauge. In chiral models, discrete symmetries are strongly constrained by anomaly cancellation conditions  \cite{Ibanez:1991hv} (see also  \cite{Banks:1991xj,Ibanez:1992ji}).
For MSSM spectra there is a short list of allowed possibilities \cite{Ibanez:1991pr} (see also \cite{Dreiner:2005rd} and references therein; see also \cite{Kurosawa:2001iq,Mohapatra:2007vd,Araki:2008ek,Lee:2010gv,Kappl:2010yu}).

Abelian $\mathbb{Z}_n$ discrete gauge symmetries can be usefully realized as discrete remnants of $U(1)$ gauge symmetries with $BF$ couplings. This structure occurs frequently for open string sector $U(1)'s$ in orientifold models (see \cite{BerasaluceGonzalez:2011wy} for a study in geometric orientifold compactifications). These models  are obtained by starting from a closed type-II 
superstring theory, modding out a world-sheet orientation reversing symmetry, and adding 
an open string sector, with the open strings ending on a brane (see \cite{Ibanez:2012zz} for review and references). 
 The Chan-Paton group of each brane is either $O(N), Sp(N)$ or $U(N)$. In the latter case there is always a $U(1)$ gauge symmetry,
 which is often anomalous. If so, the anomaly is cancelled by a Green-Schwarz mechanism involving one of the RR axions, of which 
 there are usually many. In this process the $U(1)$ gauge boson acquires a mass through a $BF$ couplings, equivalently, by absorbing a scalar field (henceforth dubbed `axion') as its longitudinal component.
 Furthermore, even non-anomalous $U(1)$'s can acquire a mass through this kind of $BF$ interaction \cite{Ibanez:2001nd}.
 
 The perturbative couplings always respect charge conservation with respect to any of these $U(1)$'s, anomalous or not, massive or massless.  For example, in brane models where baryon number and lepton number are embedded in these brane $U(1)$'s, no $B$ or $L$ violating couplings are generated perturbatively. However, massive $U(1)$'s are in general violated by non-perturbative effects, mediated by brane instantons coupling to the relevant axion field. The branes supporting the instanton may be part of the brane configuration of the model, in which case they correspond to gauge instantons, and have strength controlled by gauge couplings, hence negligible in SM-like models. More interestingly, the instantons can originate from different branes, and have strength unrelated to gauge couplings, and hence can potentially induce sizable values for the coefficients for the $U(1)$ violating operators. Examples of such 
 ``stringy instantons" have been discussed in many papers  \cite{Blumenhagen:2006xt,Ibanez:2006da,Florea:2006si,Blumenhagen:2009qh,Cvetic:2011vz,Blumenhagen:2007zk,Cvetic:2007qj,Cvetic:2007ku,Ibanez:2008my,Ibanez:2007rs}.
Such instantons may in general have a surplus of zero-modes, so that they do not contribute to the superpotential, but the violation of the symmetry arises in higher derivative terms \cite{Beasley:2004ys,Beasley:2005iu,GarciaEtxebarria:2008pi}. It is thus important to characterize the general set of $U(1)$ violating instantons in any given string model.

This is in general difficult to achieve by explicit construction. If a brane instanton is found with suitable zero-mode structure to induce a coupling,  this proves that the latter is generated;  but if no such explicit instanton can be found, it does not follow that the coupling is {\it not} generated. This is especially true in RCFT constructions, where {\it a priori} only a limited set of branes (boundary states) is available.
This is where discrete symmetries can play a useful r\^ole. It may happen that a $U(1)$ is not broken completely, but that a discrete $\mathbb{Z}_n$ symmetry remains. In that case, no instanton, and indeed no no-perturbative effect can break it, since it is an exact symmetry of the theory.
  
A second motivation to study the existence of discrete gauge symmetries is that, apart from forbidding undesirable terms, discrete symmetries may also forbid desirable ones, such as certain Yukawa couplings, or Majorana mass terms for right-handed neutrinos. Hence, the study of discrete symmetries can help one to avoid the pointless exercise of looking for instantons that cannot exist anyway, or to focus on those models which potentially allow for them. 

Finally, one may also ask a different kind of question. Rather than determining if a given model
does or does not have a certain discrete symmetry, one may ask if the occurrence of such symmetries is a generic phenomenon in string theory, or at least in certain subclasses. If the answer is positive, an appeal to discrete symmetries to avoid catastrophic couplings
becomes {\it more} credible, but then it becomes {\it less} convincing that e.g. Majorana mass terms  or perturbatively forbidden Yukawas will generically be generated by instantons.

In this respect, we note that our results on the frequency of occurrence  of discrete symmetries are not directly related to other work on the occurrence of  discrete R-symmetries in the string landscape, such as \cite{DeWolfe:2004ns,DeWolfe:2005gy,Dine:2005gz}. The latter deal with closed type-II strings (including some type-II Gepner models), without open string sectors, whereas we study symmetries originating from Chan-Paton groups of open strings. Also, the discrete symmetries discussed in the present paper act on all members of supermultiplets in the same way, and are therefore {\it not} discrete R-symmetries. Nevertheless, our discussion {\it does} include R-parity, since despite its name it can be obtained as the discrete $\mathbb{Z}_2$ subgroup of $B\!-\!L$ generated by $(-1)^{3(B-L)}$ (physically indistinguishable from the $\mathbb{Z}_2$ R-symmetry $(-1)^{3(B-L)+2S}$, since in any scattering amplitude $S$ is conserved modulo integers).

For all the reasons explained above it is  important to be able to compute in an efficient way and in a large classes of models if these orientifold $U(1)$'s are completely
or only partly broken. This is what we wish to do here for the class of Gepner orientifolds (orientifolds  of the closed type-II string theories first constructed in \cite{Gepner:1987vz}).

Gepner orientifolds models \cite{Angelantonj:1996mw,Aldazabal:2003ub,Blumenhagen:2004cg,Dijkstra:2004cc,Aldazabal:2004by,Dijkstra:2004ym,Anastasopoulos:2006da,Aldazabal:2006nz,Kiritsis:2009sf,Anastasopoulos:2010hu}
provide
access to an interesting  region of the landscape where open string configurations can be found that realize the standard model. 
The methods used are quite different than the more familiar ones used for orientifolds of orbifolds or  Calabi-Yau models. 
The basic ingredients are not branes wrapping cycles on a manifolds, but boundary states in rational conformal field theory. However
most of the techniques available in brane descriptions can be translated rather easily to rational CFT language.
For example, it
is known how to compute massless (and even massive) spectra corresponding to brane intersections, 
how to cancel disk tadpoles against crosscap tadpoles, and 
how to check if a $U(1)$ is broken by couplings to axions. 
Examples exist \cite{Dijkstra:2004ym}
where the resulting unbroken gauge group is  {\it exactly} $SU(3)\times SU(2)\times U(1)$. In this example 
all tadpoles cancel within the standard model sector, hence
there are no ``hidden sector" gauge groups, and all superfluous continuous $U(1)$'s, especially the usually problematic $B\!-\!L$ symmetry, are broken. 

In \cite{BerasaluceGonzalez:2011wy}
 it was explained how to find $U(1)$'s broken to discrete subgroups in the geometric approach. We would like
to translate this result to RCFT, or at least the special case of Gepner models. Here we immediately run into a problem. 
In the geometric setting, discrete symmetries can be read of from the axion-gauge boson couplings, which in a suitable
geometric basis are integers. The signature of a discrete symmetry is a common factor of these integers. However, in a RCFT
setting, the boundary coefficients are complex numbers, and a canonical basis in which they are integers is not readily
available. 

In this paper we solve this problem in an empirical way, by developing an algorithm that does allow us to write 
all coefficients in terms of integers. This algorithm yields a preferred set of boundary states that plays the same r\^ole as
the aforementioned geometric basis.
This involves extensive numerical computations, which give the desired result in all
cases. The fact that this is possible
calls for a deeper understanding, a principle that determines the 
basis without extensive computations.  This in its turn may provide new insights in the geometric interpretation of all these
orientifold models. However, for our present purpose the empirically determined basis does the job. We hope to return to
the underlying structure in the future.

This paper is organized as follows. In the next section we review some basic details of Gepner Models.
In section 3 we define the problem and explain how it is solved. In section 4 we explain the algorithm that 
yields the integral basis. In section 5 we give some examples. Some conclusions are offered in section 6.

\section{Gepner orientifolds}

A given Gepner model is specified by a tensor product of N=2 minimal models and a modular invariant
partition function (MIPF for short). The tensor product has total central charge 9. There are 168 ways of tensoring
minimal N=2 models so that a the required central charge 9 is obtained. They can be denoted as $(k_1,\ldots, k_N)$, where
$k_i \geq 1$ is the level of factor $i$. Usually we drop the commas and the brackets, and denote a tensor product by
a numerical string built out of the $k_i$ in ascending order. This notation turns out to be unambiguous. 
Each of the 168 Gepner models has a chiral algebra ${\cal X}$, containing the N=2 super-Virasoro algebra. This algebra
has a finite number (typically of order $10^3 \ldots 10^5$) of representations, labelled by a set of integers.  To get these representations,
one starts with all combinations of representations of the $N$ tensor factors, and subjects them to a projection to impose
world-sheet supersymmetry. In addition to the $N$ tensor factors, the four-dimensional NSR fermions are also participating in a 
non-trivial way. In addition to this, one extends the chiral algebra with a spin-1 operator which is a space-time spinor and imposes
a GSO projection. The result of these two extensions is the algebra ${\cal X}$. Here and in the following we use the simple current  \cite{Schellekens:1989am,Intriligator:1989zw}
description of MIPFs of Gepner models presented in \cite{Schellekens:1989wx}. It has the advantage that all world-sheet and space-time supersymmetry projections
can be treated on equal footing with the construction of the MIPFs, and that explicit details about the N=2 primaries can be omitted. All we need to know is how
the simple currents act on them.

The most general simple current
MIPFs for each of the 168 Gepner models can be built using the formalism developed in \cite{Kreuzer:1993tf}. This gives rise to a total of 5392
distinct MIPFs. They are characterized by a non-negative integer multiplicity matrix $M_{ij}$ indicating how often certain
left-right character combinations occur in the closed string.  The labels $i$ refer to the aforementioned set of chiral
algebra representations.
Permutations of identical N=2 factors\footnote{In the context of RCFT orientifolds, 
a naively possible origin of discrete symmetries is the permutations of identical factors. 
However, many of these permutations are broken by the MIPF, orientifold and boundary state choice. Although it is possible that cases exist where
permutation symmetries survive in the spectrum as exact discrete symmetries, we do not know any examples. In any case this is not the subject of the present paper, which as already mentioned focuses on symmetries arising from open string sector $U(1)$'s.} generate isomorphisms
between MIPFs. These have been removed, so that with a few exceptions all MIPFs are really distinct. There are just a few
cases where two or more supposedly distinct MIPFs yield apparently identical data. 

On top of this there are choices to define the orientifold quotient \cite{Fuchs:2000cm}. Taking all of these into account brings the total number of possibilities to 
about 49000. However,  some of them have zero tension, which implies that no branes can be 
added without violating the dilaton tadpole condition. After eliminating these cases we end up with $32990$ in principle
distinct orientifolds (as with the MIPFs, in practice there are always a few ``accidental" degeneracies that are apparent in the spectrum, 
but do
not have a very obvious fundamental origin. This is irrelevant in practice).

  A simple current MIPF is characterized by a discrete group ${\cal H}$ of simple currents, and a matrix of rational numbers
$X(M,J)$ defined on ${\cal H}$.  
On Riemann surfaces with boundaries each MIPF has a definite set of Ishibashi states and a corresponding set of boundary states.
The former are simply in one-to-one correspondence with the elements $M_{ii^c}$ of the multiplicity matrix, where $i^c$
denotes the two-dimensional charge conjugate of $i$. In a simple current MIPF these states are labelled by a label $m$ referring
to a representation of the chiral algebra of the tensor product, and a degeneracy label $J$. For each $m$, this degeneracy label
is the simple current in ${\cal H}$ that fixes it, {\it i.e.} $Jm=m$, with $M_{mm^c} \not=0$.  So Ishibashi states will be denoted as
$(m,J)$.

The set of boundary states that respects all the symmetries of the original chiral algebra is known to be equal to the
number of Ishibashi states \cite{Pradisi:1996yd}. They are characterized by the orbits of ${\cal H}$ on the chiral algebra representations. These
orbits can be labelled by an integer $a$ that belongs to the set of representation labels of the full  chiral algebra. An orbit
is a set of representation labels related by the action of ${\cal H}$. For the boundary label we choose one representative from this set. 
Also in this case there may be degeneracies, which occur if the ${\cal H}$-action has fixed points. 
The degeneracy labels can be conveniently chosen as the discrete group character $\psi$ of certain
subgroup (called the ``central stabilizer") ${\cal C}_a$ of the stabilizer ${\cal S}_a$ of $a$ (the stabilizer is the subgroup  
of ${\cal H}$ of that fixes a representation $a$).  The boundary labels are then $[a,\psi_a]$. 
Note that the set of characters depends
on the boundary label. If the central stabilizer is a discrete group with  $|\cC_a|$ elements, than there exists exactly
$|\cC_a|$ distinct characters (complex functions on ${\cal C}_a$ that respect the group property).

Now we have two sets $(m,J)$ and $[a,\psi_a]$ of Ishibashi and boundary labels. These can be shown to be of equal
size, although this is not manifest. On this basis we now define boundary reflection coefficients \cite{Fuchs:2000cm}
\beq \label{eq:BRC}
R_{[a,\psi_a](m,J)}  =  \sqrt{\frac{|\cH|}{|\cC_a||\cS_a|}}
 \psi_a^*(J) S^J_{am}
\eeq
Here $S^J_{am}$ is a matrix element of the modular  transformation matrix of a certain algebra associated with
the original chiral algebra and the current $J$ \cite{Schellekens:1989uf,Fuchs:1995zr}. If $J$ is the identity, $S^J$ is equal to the modular transformation 
matrix of the chiral algebra ${\cal X}$. If $J$ is not the identity,  $S^J$ is the modular transformation matrix of another
algebra, usually, but not always, related to some other conformal field theory.
All these matrices are explicitly known and are, in general, complex numbers.
In the prefactor $|\cH|$, etc,  denotes the number of elements of the corresponding discrete group.

The boundary coefficients are independent of the orientifold choice. The latter enters the discussion in
two ways. First of all the unoriented annulus coefficients have the form \cite{Fuchs:2000cm}
\beq
A^{i}_{[a,\psi_a][b,\psi_b]}  =  \sum_{m,J,J'}
	\frac{ S^i_{~m}R_{[a,\psi_a](m,J)} g^{\Omega,m}_{J,J'} R_{[b,\psi_b](m,J')} }{S_{0m}} \label{Acoef} \\
\eeq

Here $g^{\Omega,m}_{J,J'}$ is an orientation-dependent metric on the space of Ishibashi states; $\Omega$ denotes
the orientifold choice. In general, $g^{\Omega,m}_{J,J'}$ is a block-diagonal matrix in the label $m$, which can  act
non-trivially in the degeneracy spaces for each $m$.
One could in principle take the square root of this metric and absorb it into the
boundary coefficients, which then become orientation dependent. However, it is both physically more appealing and also
more convenient to have orientifold-independent boundary coefficients. The final results will not be affected by this convention.

To make the notation a bit less cumbersome we will use in the following a single letter ``$a$" instead of the combination
$[a,\psi_a]$ to denote boundaries. The fixed point splitting of the boundary labels does not really play a r\^ole in what follows. 

The annulus coefficients appear in the expression for the
oriented annulus as
\beqa
A  = \sum_{a,b}  N_a N_b     \sum_i A^{i}_{ab} \ \chi^i(\tau/2) \ ,
\label{annulus}
\eeqa
where   $N_a$, $N_b$ are the Chan-Paton multiplicities, and $\chi^i$ are the Virasoro characters of the
chiral algebra ${\cal X}$.

The second way the orientifold choice matters is in
boundary conjugation. This is defined as follows
\beq
A^{0}_{ab} = \left\{ \begin{array}{ll}
					1 & \hbox{if}\ b = a^c \\
					0 & \hbox{otherwise}
					\end{array}
\right.
\eeq
Clearly  the dependence of boundary conjugation on orientation can be traced back to the Ishibashi metric $g^{\Omega,m}_{J,J'}$, so
that in the end all dependence on orientation can be traced back to this quantity.

\section{Axion couplings}

\subsection{Discrete $\mathbb{Z}_N$ symmetries from open string $U(1)$'s}

The key to understand the appearance of discrete $\mathbb{Z}_N$ gauge symmetries from open string $U(1)$'s are the coefficients $R_{am}$, which determine the $BF$ couplings.

Consider a 4d string model, with a set of branes labelled with $a$ and their orientifold images $a^c$, with $BF$ couplings to a set of RR 2-forms $B_m$
\beqa
\sum_{a,m} N_a \, V_{am}\, B_m\wedge F_a
\eeqa
Here $V_{am}=R_{am}-R_{a^cm}$, with the relative minus sign arising because the physical $U(1)$ gauge boson is the difference of those supported on the brane and its orientifold image. 

Consider now a linear combination\footnote{It is useful to maintain the convenient normalization that $U(1)$'s have minimal charge 1; this requires the $x_a$ to be integer, with gcd$(x_a)=1$.\label{normalization-u1}}  $\sum_a x_a Y_a$ of the $U(1)$ generators $Y_a$ of brane $a$. 
Its $BF$ couplings are
\beqa
\sum_m \big(\, {\textstyle\sum}_a x_a N_a V_{am}\, \big) B_m \wedge F
\label{bf}
\eeqa
It thus remains massless if and only if
\begin{equation}\label{UMass}
\sum_a  x_a N_a (R_{am}-R_{a^cm})  = 0\   \hbox{for all}\ m.
\end{equation}
In general, the set of massless $U(1)$'s correspond to the space of zero eigenvectors $x_a$
of the non-symmetric matrix $M_{am}=N_a (R_{am}-R_{a^cm})$. 

Massive $U(1)$'s are broken by brane instantons coupling to the axion RR scalars $\phi_m$ dual to the 2-forms. With a suitable normalization, the amplitudes go like $e^{-2\pi i\phi_m}$, and the axions have an identification $\phi_m\simeq \phi_m+1$. It is useful to introduce the dual description of (\ref{bf})  in terms of $\phi_m$.
The relevant lagrangian is
\beqa
\sum_m \left[ \, \partial_\mu\phi_m\, -\, (\textstyle{\sum}_a x_a N_aV_{am})\,A_\mu \right]^2
\label{axionlag}
\eeqa
where the $U(1)$ is normalized such that the minimal charge is 1. Under $U(1)$ transformations, 
\beqa
A_\mu\to A_\mu +\partial_\mu \lambda \quad ; \quad \phi_m\to \phi_m+ (\textstyle{\sum}_a x_a N_aV_{am})\lambda
\label{gauge-transf}
\eeqa
Instanton amplitudes transform as
\beqa
e^{-2\pi i\phi_m} \to e^{-2\pi i\phi_m} \exp[-2\pi i (\textstyle{\sum}_a x_a N_aV_{am})\lambda]
\eeqa
and this transformation is cancelled by the insertion of an operator in the charged matter, with total charge $(\textstyle{\sum}_a x_a N_aV_{am})$. This quantity therefore measures the amount of $U(1)$ violation.

Thus the condition for a discrete $\mathbb{Z}_N$ remnant of the $U(1)$ is therefore 
\begin{equation} \label{Null}
\sum_a  x_a N_a (R_{am}-R_{a^cm})  = 0 \ {\rm mod} \ N \ {\mbox{ for all $m$} }.
\end{equation}
This $\mathbb{Z}_N$ is  an exact gauge symmetry. The result also follows from the analogy of (\ref{axionlag}) with the Higgsing of the $U(1)$ by a charge $N$ scalar (whose phase is played by a suitable linear combination of the RR scalars). 

In other words, to find $\mathbb{Z}_N$ discrete symmetries, we should look for zero eigenvectors modulo $N$
of the matrix $M_{am}$, in the convenient normalization used above. In a geometric setting, they can be made integer by choosing a suitable basis for the axions, in terms of basic 3-cycles on the compactification manifold (in type-IIA language), see \cite{BerasaluceGonzalez:2011wy}. But this notion (and so the automatic appearance of the convenient normalization) is not readily available for Gepner models, although in some very special cases (the quintic Calabi-Yau) similar bases have been discussed \cite{Brunner:2004zd}. A specialized geometric discussion for each separate case is not likely to get us to the desired result, since we will have to deal with all possible simple current MIPFs of the 168 Gepner models, and all their orientifolds, a total of 32990 distinct possibilities with non-vanishing orientifold tension. 

The general formula for the boundary coefficients in such a CFT takes the form  (\ref{eq:BRC}).
Here $S^J_{am}$ is a modular transformation matrix of a conformal field theory, and $\psi$ is a phase, and the pre-factor
is a square root of a rational number. Neither of these factors are integers. Indeed, in general these boundary coefficients are complex. It is not clear how to even define condition (\ref{Null}). The key towards the resolution, is to search for a basic set of instanton branes, whose boundary coefficients thus define the axion periodicities; this then allows to effectively move onto the convenient normalization in which coefficients are integers, whose gcd then gives the order of the discrete symmetry.

\subsection{Axions  in RCFT and basis of boundary states}

In this section we describe the structure of axions in RCFT orientifolds, and explain the relevance of the above mentioned basis of boundary states in RCFT terms.

An Ishibashi state $(m,J)$ contains an axion if the representation $m$ contains a massless space-time spinor. 
The ground state may contain $N_L$ left-handed and $N_R$ right-handed massless spinors. In the closed string
one gets the square of the character multiplied by the multiplicity matrix $M_{ij}$. This is then subject to the Klein
bottle projection if $i=j$ (since the Ishibashi states correspond to $i=j^c$, the Klein bottle
projection can only affect self-conjugate Ishibashi states with $N_L=N_R$;  for a more detailed discussion see the appendix).
However all these multiplicities are ignored in the following, because they all have the same boundary coefficients. 
Each representation with $N_L+N_R > 0$ 
is counted as one axion, regardless of the values of these numbers and Klein bottle projected closed string
multiplicity  and even if  the Witten index $N_L-N_R$ vanishes. 
Note however that each degeneracy label $J$ is counted once as a different axion, since the axion coefficients depend on $J$. 

We introduce the following notation for the relevant coefficients.
\beq
V_{a\nu} \equiv R_{a(m,J)} - R_{a^c (m,J)} \ ,
\eeq
where $a=1,\ldots,N_B$, the number of complex boundary pairs.
The second label $\nu$ identifies
contributing axions according to the rules stated above. We will discuss the precise range of the label below. 
Charge conjugation is defined by means of the orientifold choice. 
We regard these objects as a complex matrix with columns labelled by $\nu$ and rows by $a$.
Note that self-conjugate boundaries do not contribute, and that each complex pair contribute one row, so that there is a one-to-one
correspondence between the rows and all possible $U(1)$ factors in the open string spectrum.

As stated above, these coefficients $V_{a\nu}$ are in general complex. However, since they are coefficients of $BF$ couplings, they should be real (morally, up to phase redefinitions of the RR fields); in other words, they can be made real by an $a$-independent but $\nu$-dependent phase rotation (as can be shown directly in explicit models, see next section), as we assume in the following.

Note however that, since they axion periodicities had not been fixed to unity, they are not integers or rational numbers. However we will demonstrate in the next section
that there exists at least one choice of boundaries $c(\mu)$ 
such that the following relation holds
\beq\label{Basis}
V_{a\nu}= \sum_{\mu=1}^{N_A} Q_{a\mu} V_{c(\mu)\nu} \ ,\ \ \  Q_{a\mu} \in \mathbb{Z}
\eeq
Here $c$ is a map from the set of axion labels into the set of boundary labels which assigns a different
boundary label to each axion label: $c(\mu) \not=c(\nu)$ if $\mu \not= \nu$. The number of basis vectors $N_A$ is
equal to the number of independent columns of $V_{a\nu}$ (it turns out to be sufficient to remove vanishing and identical
columns, more complicated dependencies do not occur). Consequently, the labels $\nu$ cannot be expressed unambiguously
in terms of the original transverse channel labels $(m,J)$: there may be more than one $(m,J)$ corresponding to any given $\nu$.
More details will be given in the next section.

If (\ref{Basis}) can indeed be realized, it defines a basis in the space
of all complex boundaries such that all other boundaries can be expanded in that basis with integer coefficients. In this way
we obtain a lattice of charges, so that each boundary state corresponds to a point on that lattice. In general we expect that
this basis is not unique, just as the basis of a lattice is not unique. Note however that not every lattice point is occupied. This
is obvious because there is only a finite number of boundary states and an infinite number of lattice points, but also near the 
origin there are in general unoccupied sites. This implies that not every lattice basis can be realized in terms of boundary states.

The basic boundary states defines a set of `smallest instantons' (at least in the RCFT realm), whose couplings to the axions {\em define} the axion periodicities. The quantities $Q_{a\mu}$ thus correspond to the coefficients of the BF couplings in the desired normalization in which the axions have unit periodicity, and can therefore be used to look for the discrete $\mathbb{Z}_N$ symmetry. Namely a $U(1)$ integer linear combination $Y=\sum_a x_a Y_a$ (with the conventions in footnote \ref{normalization-u1}) has an unbroken $\mathbb{Z}_N$ subgroup if it satisfies the condition 
\beq\label{Condition}
\sum_a x_a N_a Q_{a\mu} =0\mod N
\eeq 

There is an alternative description of the physical relevance of the basis, which instead of leaning on the axion periodicities, is based on expressing the amount of instanton $U(1)$ violation in terms of the basic instantons, as follows (both viewpoints are clearly related since (\ref{gauge-transf}) links $U(1)$ gauge transformations and axion shifts). As described in \cite{Blumenhagen:2006xt,Ibanez:2006da}, the amount $I_b(a)$ of $U(1)_a$ violation  by an instanton supported on a brane $b$ is given by the net number of charged fermion zero modes arising from massless open strings stretching between both boundaries. In the RCFT setup, and accounting for orientifold images, we have a combination of the annulus coefficients (\ref{annulus})
\beqa
I_b(a)\,=\,N_a \sum_i w_i (A^i_{~ba} - A^i_{~ba^c} )
\eeqa
where $w_i$ is the Witten index in the open string sector, which effectively extracts the net chiral contribution. Using (\ref{Acoef}) we have
\begin{align}
I_b(a) = N_a\sum_i w_i \sum_{m,J',J} \left[{ S_{im} R_{b(m,J')} g^{\Omega,m}_{J'J}} \over{ S_{0m}}\right] (R_{a(m,J)} -R_{a^c(m,J)})
\label{masslessu3}
\end{align}
Note that these quantities are integer, and moreover can be defined for any boundary states $a,b$, regardless of whether $a$ actually realizes a $U(1)$ symmetry in the model or not.
Decomposing the boundary coefficients using (\ref{Basis}), and reconstructing back to annulus amplitudes, we obtain
\beq
I_b(a) = \sum_{\mu} N_a Q_{a\mu} I_b(c(\mu))
\label{basic-fmla}
\eeq
Here $I_b(c(\mu))$ are formally defined as in (\ref{masslessu3}); in physical terms, they are integers measuring the violation by the instanton brane $b$ of a {\it putative} $U(1)$ carried by brane $c(\mu)$ (which need not support an actual $U(1)$ of the model). For a $U(1)$ linear combination $Y=\sum_ax_aY_a$ (with the conventions in footnote \ref{normalization-u1}), the charge violation by an instanton brane $b$ is 
\beqa
I_b(x)\, =\, \sum_a x_a I_b(a)\, =\,\sum_{\mu} \big(\, \sum_a x_a  N_a Q_{a\mu}\, \big) I_b(c(\mu))
\eeqa
Since $I_{b}(c(\mu))$ are integer, if the coefficients $\sum_a x_a  N_a Q_{a\mu}$ have a common factor $N$, all instantons violate $U(1)$ charge in multiples of $N$, so that a discrete $\mathbb{Z}_N$ subgroup remains unbroken.  Hence we recover condition (\ref{Condition}) for the existence of a discrete $\mathbb{Z}_N$ symmetry.

Although this derivation exploited the RCFT formulas, eq. (\ref{basic-fmla}) makes full physical sense even for non-RCFT instantons. This strongly supports that the result holds for any instanton $b$, and therefore that the proposed condition (\ref{Condition}) is correct in general. Still, it is possible that the basic quantities $I_{b}(c(\mu))$ already have a common factor. If they do not, we will get a $\mathbb{Z}_N$ discrete symmetry, as read off from the coefficients $Q_{a\mu}$; otherwise, we can only get more discrete symmetries than naively expected. We believe this possibility to be fairly unlikely. The fact that we were able to find an integral lattice of charges in all cases strongly suggests  that (\ref{Condition}) identifies the discrete symmetries correctly. 
\section{Finding an integral basis}

We will now explain a method that turns out to be very effective to find the integral basis described above.

Our starting point is the matrix $V_{a\nu}$, where rows $a$ label boundary states and columns $\nu$ label axion fields. First we will normalize the coefficients $V_{a\nu}$ in a convenient way. In their raw form, these coefficients are not even relatively real. However, on already explained physical grounds, they can be made real with $a$ independent phase redefinitions, which are duly accounted in the following normalization. Consider in each column the first non-vanishing entry, starting at the top. If there is such an entry, divide all entries in the column by it, so that the top entry is equal to 1. If there are any columns that are completely zero, we discard them, since they describe decoupled axions; also, if two columns are identical, we keep only one of the two, since there is a decoupled linear combination of axions. This procedure only eliminates vanishing or identical columns. This is in general not sufficient to ensure that
the columns are linearly independent, although this turns out to be the case in practice in the whole class of models. We call the dimension of this axion space $N_A$. As expected, it turns out that after normalizing the top entry of each column to 1, all entries in the matrix become real numbers.

This normalization removes some convention-dependent factors in the boundary coefficients. For instance, as mentioned below (\ref{Acoef}), we could have defined the boundary coefficients differently by absorbing the square root of $g^{\Omega,m}_{J,J'}$ in them; this is conveniently done by choosing a basis in degeneracy space so that $g^{\Omega,m}_{J,J'}$ is diagonal, so it can be absorbed into the boundary coefficients by multiplying each column by a certain complex factor. The normalization procedure discussed in the foregoing paragraph removes any possible dependence on such conventions.


Note that this normalization procedure depends on the way the boundaries are ordered. This ordering is not just
arbitrary, because it descends from the ordering of the representations of the chiral ${\cal X}$, but the ordering is not
in any way canonical either. Roughly speaking, it has the property that if $i > j$, then $S_{0i} > S_{0j}$, but even that
ordering is not strict. 
However, the final result will not depend on this normalization procedure.


In this way we now obtain a real matrix $V_{a\mu}$, where $a$ labels boundaries and $\mu$ the reduced set of axionic Ishibashi states.
Now we consider the inner product matrix 
\beq\label{InnerProd}
N_{ab}=\sum_{\mu} V_{a\mu}V_{b\mu} \equiv V_{a}\cdot V_{b}
\eeq
Upon explicit computation, it turns out that this is a rational matrix in all models, even though the coefficients $V$ are real, and in general not rational.
Note that if we renormalize an entire column by $\sqrt{q}, q \in \mathbb{Z}$ this does not affect the rationality.
However, it is not uncommon to encounter other irrational numbers such as $p+\sqrt{q}$ and sine and cosines
of rational multiples of $\pi$. It is therefore far from obvious that the rationality of $V$ will persist if we
order the boundaries differently, thus obtaining a different normalization prescription. 
However, it is an
empirical fact that in all 32990 cases of different MIPFs and orientifolds all these numbers $N_{ab}$ come out rational, and this will turn out to
be a very fortunate outcome.

Based on the intuitions in earlier sections, the hope is to find a basis in the space of Ishibashi states such that
all coefficients $V_{a\mu}$ are transformed into integers, {\it i.e.}
find a real and invertible matrix $R$ such that
\beq
Q_{a\nu} =\sum_{\mu} V_{a\mu}R_{\mu\nu}   \in  \mathbb{Z}  \label{Qdef}
\eeq
If such a basis exists, the coefficients $V_{a\mu}$ can be written as  
\beq
V_{a\nu} =\sum_{\mu} Q_{a\mu}R^{-1}_{\mu\nu}  
\eeq
We may think of the  matrix $R^{-1}_{\mu\nu}$ as a set of basis vectors $B_{\nu}^{(\mu)}$ labelled by $\mu$, and then what we are
looking for is a set of basis vectors in terms of which all vectors $V_{a\nu}$ have integer expansions. In other words, all vectors $V_{a\nu}$ lie on the lattice spanned by the basis vectors.  
If we express the inner products $N_{ab}$ in (\ref{InnerProd}) in terms of the basis vectors we get
\beq
N_{ab}=\sum_{\mu} \sum_{\nu} Q_{a\mu}Q_{b\nu}\sum_{\rho} R^{-1}_{\mu\rho}R^{-1}_{\nu\rho}=
\sum_{\mu} \sum_{\nu} Q_{a\mu}Q_{b\nu}\ B^{\mu} \cdot B^{\nu}
\eeq
This tells us that if
the basis vectors have integer (or rational) inner products, then integrality (rationality) of {\it all} $N_{ab}$ follows
automatically.

It is then natural to conjecture that the basis vectors might themselves be chosen as a subset  of the boundary 
vectors $V_{a\mu}$.
A necessary condition is that we should be able to find $N_A$ independent vectors $V_{a\mu}$. Here it is important that
the $N_A$ columns are linearly independent, as explained above. 
A basis of this kind is defined by a map
$c(\mu)$ from 
the set of axion labels to the set of boundaries, and we write
\beq
R^{-1}_{\mu\nu}=B_{\nu}^{(\mu)}=V_{c(\mu)\nu}
\eeq

After inverting this matrix
we can compute the charges using (\ref{Qdef}). The fact that all $N_{ab}$ are rational guarantees that the
charges are rational. But we can do better than that.
Suppose some boundary vector $W$ has the following expansion in terms of the basis
\beq 
\label{FraqExp}
W_{\nu} =\sum_{\mu} Q_{\mu}V_{c(\mu)\nu}  = \sum_{\mu} \frac{p_{\mu}}{q_{\mu}}V_{c(\mu)\nu} \ ,
\eeq
where $p_{\mu}$ and $q_{\mu}$ are relative prime.
Now suppose there is one $\mu$, denoted $\hat \mu$, so that $p_{\hat\mu}=1$. We may then bring the corresponding term to the left, $W_{\nu}$ to the right and
multiply by $-q_{\hat \mu}$. 
Then we get
\beq
V_{c(\hat \mu)\nu} =  \sum_{\mu, \mu \not= \hat\mu}    -\frac{p_{\mu}q_{\hat\mu}}{q_{\mu}}V_{c(\mu)\nu} + q_{\hat\mu} W_{\nu}
\eeq
Now we may remove $V_{c(\hat \mu)}$ from the basis and replacing it by $W$, thus defining a new map, $\hat c(\nu)$.
The advantage is that now one of the charges has changed from $1/q_{\hat\mu}$ to $q_{\hat\mu}$. Furthermore, if 
$q_{\mu}$ and $q_{\hat \mu}$ have common factors, the remaining denominators are reduced (in the majority of cases all
denominators in (\ref{FraqExp}) are in fact equal to $q_{\hat\mu}$, so that all coefficients become integer). 

Now we iterate this process: compute all charges of the boundary vectors with respect to all basis vectors, and as soon
as we encounter one with charge $1/q$, we interchange the corresponding basis vector and boundary vector. Note that in
every step the the determinant of the inner product matrix of the basis vectors (which is the square of the volume of the
unit cell of the lattice) is reduced\footnote{Proof: Consider  the lattice spanned by the $N_A-1$ vectors $V_{c(\mu)}$, with $V_{c(\hat\mu)}$ removed. The volume of the full unit cell is the volume
of the unit cell in this  $N_A-1$ dimensional sub-lattice, times the length of $V_{c(\hat\mu)}$ times $\rm sin\ \theta$, where $\theta$ is
the angle between $V_{c(\hat\mu)}$ and the plane of the sub-lattice. The new vector $W$ can be decomposed
in a component along $V_{c(\hat\mu)}$ and a component in the plane of the sub-lattice. The component of $W$ along 
$V_{c(\hat\mu)}$ has a length $1/q$ of $V_{c(\hat\mu)}$, and the projection on the sub-lattice is irrelevant for
the computation of the volume. Hence the volume decreases by $1/q$.}  by a factor $q^2\,$. This means that the procedure must end after a finite number of steps.

The only way  the procedure can fail is if no charge $1/q$ can be found.
A simple example demonstrating such a failure is a one-axion case with just two boundary vectors $v_1=(2)$ and $v_2=(3)$. 
There are two possible bases, and the only charges we encounter are either $\frac23$ or $\frac32$.  This situation
never occurs for any of the 32990 Gepner orientifolds. However, 
it may also happen that an integer basis exists, but that the algorithm converges to an incorrect basis. We did indeed encounter
just three cases where we ended up with a basis with respect to which all charges are either integer, or half-integer, with values
$q/2$, $|q| \geq 3$. Then no further progress is possible. These three cases could be handled by reordering the initial set of boundaries,
so that the algorithm converges to a different set. For all 32990 orientifolds a maximum of 19 iterations was necessary to reach an integer basis.

Note that all charges are defined in terms of boundary vectors, as announced in (\ref{Basis}), through
\beq\label{Main}
V_{a\nu}= \sum_{\mu} Q_{a\mu} V_{c(\mu)\nu} 
 \eeq
 so that the original basis in which the boundary vectors are expressed is irrelevant. In particular, the
 unusual normalization procedure of the columns drops out between the left- and righthand side. However,
 this normalization procedure lead to rational inner products, which was a great convenience for obtaining the result.

A final comment is that, although the existence of a basis is expected on general grounds (e.g. extrapolation from the geometric regime, or ultimately, from brane charge quantization), it is very remarkable that in the present setup the basis is realized in terms of RCFT boundary states. In particular, this implies that at the Gepner point all the basis vectors are mutually supersymmetric, a very special configuration reminiscent of fractional branes at singularities. It is possible that work along the lines of \cite{Diaconescu:2000ec}, realizing rational boundary states as fractional branes in LG orbifolds, further clarifies the nature of the above basis beyond the brute force construction.

\subsection{Example: The Quintic}

In order to explain how the algorithm explained above works in practice, we will consider here the quintic, a well studied case
in the comparison between Gepner models and Calabi-Yau compactifications. An integral basis for tadpole charges for this case
was presented in \cite{Brunner:2004zd}. Here we need only a subset of those charges, since we are only interested in
the ``imaginary" boundary combinations $a-a^c$.  However, the results of  \cite{Brunner:2004zd} are not sufficiently explicit to
make a direct comparison possible, and furthermore there will in any case be a basis dependence. 

The quintic Calabi-Yau has Hodge numbers $(h_{11},h_{21})=(1,101)$, and can be obtained from the Gepner model $(3,3,3,3,3)$.
As a RCFT, this has 4000 boundary states. The total number of independent axions with $R_{a(m,J)} - R_{a^c(m,J)}$ couplings turns out to 100.
Of the 4000 boundary states, 32 have a Chan-Paton group $SO(N)$, and these do not couple to these axions. The remaining 
3968 boundaries are pairwise related by conjugation. Hence we end up with a total of 1984 vectors $V_{a\nu}$,  with $\nu=1,\ldots,100$.
We normalize them in the way explained above.

In order to have the best possible chance of finding the basis we first order the 1984 vectors in a convenient way. One would naively expect the
basis vectors to have the smallest norm, so we order the 1984 $V_{a\nu}$ according to increasing norm, respecting the original CFT ordering
in case of degeneracies. Then we select the first 100 independent vectors out of this set. It turns out that the first 46 are independent, and
then we have to go up to number 200 to complete the set. Now we compute the charges of all 1984 vectors with respect to this basis. To do so,
we start with (\ref{Main}) and contract both sides with the would-be basis vectors $V_{c(\mu)\nu}$. In this way the coefficients on both sides of the
equation are related to elements of the matrix $N_{ab}$, eqn (\ref{InnerProd}), which are rational. The definition of the charges now becomes
\beq\label{MainTwo}
N_{ac(\nu)}= \sum_{\mu} Q_{a\mu} N_{c(\mu)c(\nu)} 
 \eeq
We now invert the rational matrix $N_{c(\mu)c(\nu)}$. In this case it is a $100 \times 100$ matrix, which can be inverted exactly on a computer using
unlimited size integer numerators and denominators. In this way we can avoid accuracy problems with real numbers. This is essential, because in the most
difficult case we have to deal with a $480 \times 480$ matrix.
Using the inverse we now compute
the charges. Obviously the charges of the basis vectors themselves are integers by construction, $Q_{c(\nu)\mu}=\delta_{\nu\mu}$, but this leaves 1884 
non-trivial vectors to be checked, each with 100 charges. In this example, boundaries $1,\ldots, 46$ are in the basis, boundary 47 is not, but turns out to have integral charges, but boundary
48 has charge $\frac12$ with respect to the second basis vector. 
So following the algorithm explained above we now take boundary 48 as our second basis vector. 
We recompute the inverse of the new matrix $N_{c'(\mu)c'(\nu)}$, where $c'$ denote the new basis choice.
In the next iteration boundary 53 turns out to have charge $\frac12$ with respect to basis vector 15. So we put it in the basis and try again. Now boundary 104 turns out
to gave charge $\frac32$ with respect to basis vector 6, a charge that is unsuitable, but it has charge $-\frac12$ with respect to basis vector  7. After putting boundary 104
in the basis instead of this vector, we find that all 1984 boundaries now have integer charges.

\section{Results}

In \cite{Anastasopoulos:2006da}  19345 distinct chiral classes of brane configurations were found that agree with the
standard model chirally\rlap.\footnote{All spectra are available online at
{\tt http://www.nikhef.nl/$\sim$/t58/Site/String\_Spectra.html}. They were assigned a unique number to identify them, and to
which we will refer henceforth. To examine an explicit sample of a spectrum in one of the 19345 classes, follow the instructions given
on this webpage.} These spectra are distinguished by comparing them modulo non-chiral (vector-like) matter. They all contain a group
$SU(3)\times SU(2)\times U(1)$, and all matter that is chiral with respect to that group must form {\it exactly} three families of quarks and
charged leptons. They are distinguished by their complete Chan-Paton group, the chiral matter with respect to that group, and the massless
vector bosons that exists in addition to $Y$. These Chan-Paton {\it chiral} spectra may contain matter that is $SU(3)\times SU(2)\times U(1)$ non-chiral,
such as Higgs pairs and right-handed neutrinos, as well as less desirable vector-like particles. Individual models in each class differ in their
fully {\it non-chiral} spectra, {\it i.e.} in matter that is vector-like with respect to the full, unbroken Chan-Paton gauge group.
We can now in principle investigate all of them for the presence of discrete symmetries. Note that
in each class there are many explicit realizations, which were collected by examining a subset of the
32990 non-zero tension orientifolds. This subset was determined by limiting, for
purely practical reasons, the total number of boundary states to 1750. 

The presence of discrete symmetries is not a property of the entire class, but must be examined for each class member separately.
In fact, according to the search philosophy of  \cite{Anastasopoulos:2006da} it would be natural to split these classes into
subsets with definite discrete symmetries. At present, only the presence of additional $U(1)$ bosons, in other words, the constraint $\sum_a n_a N_a Q_{a\mu}=0$,
 is used to distinguish classes.
 A natural refinement would be to distinguish brane configurations on the basis
of the constraint $\sum_a n_a N_a Q_{a\mu}=0 \mod N$. 

Since this refinement was not taken into account we have to examine classes of interest {\it a posteriori}. We did not do that for all
19345 classes, but limited ourselves to classes with Chan-Paton groups $U(3)\times U(2)\times U(1)\times U(1)$ or 
$U(3)\times Sp(2)\times U(1)\times U(1)$ (UUUU and USUU for short) with all chiral matter in bi-fundamentals, and 
$SU(5)$ GUT models with Chan-Paton group 
$U(5)\times O(1)$ ,  $U(5)\times U(1)$ ,  $U(5)\times U(1)\times O(1)$  or  $U(5)\times U(1)\times U(1)$.

\subsection{UUUU and USUU spectra}

The majority of the spectra in this class
are of the ``Madrid" type \cite{Ibanez:2001nd}. In this configuration all $U$ and $D$ antiquark open strings have one end on the $U(3)$ brane and the other on
one of the $U(1)$ branes (the {\bf c}-brane, conventionally\footnote{We use boldface subscripts to refer to 
one of the four standard model branes.}), whereas the charged leptons and neutrinos have their endpoints on
the {\bf c}-brane and the {\bf d}-brane. There is a more exotic, but also more problematic possibility of having some of the anti-quarks end 
on the  {\bf c}-brane and some on the {\bf d}-brane. This class was investigated in \cite{Kiritsis:2008ry}. In all but one of these models the
standard model $Y$ charge is given by 
\beq\label{YYY}
Y=\frac16 Y_{\bf a} - \frac12 Y_{\bf c} + \frac 12 Y_{\bf d} \ ,
\eeq
where $Y_x$ is the $U(1)$ generator of brane ${x}$. The signs are convention dependent, and are chosen differently in some
of the literature. There is one model in the database (Nr. 13395) with a $U(3)\times U(2)\times U(1)\times U(1)$ Chan-Paton group and 
an unconventional $Y$ charge $Y=-\frac13 Y_{\bf a} + \frac12 Y_{\bf b} +  Y_{\bf d}$. There are just eight samples of this particular model,
and none of them had discrete symmetries, so we will not consider it here.

In table \ref{Madrid} we summarize all 24 models with the hypercharge embedding chosen as in (\ref{YYY}). The 
horizontal lines separate the USUU and the UUUU models, and within these sets they separate  the standard Madrid
models (with perturbative lepton number conservation) from the non-standard ones. In the columns labelled ${\bf xy}$ etc. we list the
chiral intersection number $I_{{\bf xy}}$. If this number is negative, this implies that both endpoint branes must be conjugated. Note that in addition there
may always be non-chiral matter, which is ignored in the definition of a chiral class.
Within the USUU class, 
there is only one Madrid model possible, but in the UUUU class there are several, depending on the 
choice of the $U(2)$ representation $(2)$ or $(2^*)$ used for the standard model matter. Furthermore the Higgs bosons
can be chiral with respect to $U(2)_{\bf b}$, and hence contribute to $U(2)$ anomaly cancellation. No restriction was imposed on
the number of such ``chiral" Higgs bosons. Even if the number is zero, Higgs bosons can still occur as $U(2)$-non-chiral
particles in the spectrum.

For  each model, there
is a possibility of having just a single massless $U(1)$ boson, $Y$, or two, $Y$ and $B\!-\!L$. The latter possibility is much
more common than the second, and for some configurations the first option was not realized at all in the set of Gepner
orientifolds explored in \cite{Anastasopoulos:2006da}. In models that are not strictly of the Madrid type there is usually just
a single non-anomalous $U(1)$ and hence only $Y$ is gauged. In models 7488 and 13015 however, there is an additional
anomaly free $U(1)$, namely $Q_{\bf b} + 2 Q_{{\bf c}}$, which is unbroken in the second class. Interestingly, the unbroken 
case occurred less often than the broken case, just the other way around as for Madrid models\rlap.\footnote{Since a massless $U(1)$ requires the
set of axion charges to have a null vector, one would generally expect spectra with fewer massless $U(1)$'s to occur more frequently. Therefore Madrid
models are probably the exception and not the rule. Presumably this is due to the fact that in Madrid models the  {\bf a} and {\bf d} play a symmetric role, so that
often their axion couplings are identical.  This implies a massless $B\!-\!L$.  In particular this is true for a large subclass related to Pati-Salam models.}
There are several conventions one  can choose to represent these spectra. One may conjugate the ${\bf b}$ brane, which
does not participate in $Y$. Furthermore, one can conjugate the ${\bf d}$ brane and simultaneously interchange the assignment
of right-handed neutrinos and charged leptons, thus keeping $Y$ unchanged. The same can be done with the ${\bf c}$ brane. 
Our conventions are such that in the multiplicities of the fields are as much as possible positive. 
In comparison with
\cite{BerasaluceGonzalez:2011wy}, formula (3.14), the first option corresponds to numbers 10551 and 1352, after conjugation
of   ${\bf d}$. The second option, shown in parentheses in \cite{BerasaluceGonzalez:2011wy}, corresponds
to 12106 and 7976.

Note that in non-Madrid models lepton number is not defined in terms of brane charges, and hence there is an intrinsic confusion
between $L$, $H_d$ and the conjugate of $H_u$. These spectra are accepted as standard-model-like on the basis of a correct
count of the net number of $(1,2,-\frac12)$ representations, {\it i.e.} $I_{{\bf bc}^*}+I_{{\bf b}^*{\bf c}^*}+I_{{\bf bd}^*}+I_{{\bf b}^*{\bf d}^*}=3$.
Exactly which particles should be identified as Higgses or lepton doublets depends on the superpotential couplings and on the direction
of the Higgs vev in the space of these fields. This requires additional assumptions. In \cite{Kiritsis:2008ry} this was discussed for some
models in class 14062.

How generic are discrete symmetries in these 24 models? The total number of 24 classes splits into 18 with a unitary weak group and
6 with a symplectic one. In total, these classes contain 962958 brane configurations. 
The  discrete symmetries we have encountered are
$\mathbb{Z}_2$ and $\mathbb{Z}_3$ and occur only for 6 of these 24 models. We only consider a discrete symmetry if it is not contained in the continuous symmetries. 
With $Y$ of the form (\ref{YYY}), there is automatically a null-vector $n^a$ of the form $(1,0,-3,3)$. Reduced modulo 3 this yields 
$(1,0,0,0)$, and hence there is automatically a $\mathbb{Z}_3$ symmetry corresponding to $n^a=(1,0,0,0)$. This is of no interest, since
it is just the $SU(3)$ color selection rule requiring that all amplitudes be color (and hence triality) singlets. Not surprisingly, this also follows from
$Y$-charge conservation. Similarly, in models with a $U(2)$ factor there is a $\mathbb{Z}_2$ null vector $n^a=(0,1,0,0)$. This does not
follow from any continuous charge conservation, but it is equally uninteresting, since it just imposes $SU(2)$ duality. Both of these discrete symmetries are a direct consequence of the factors $N_a$ in (\ref{Condition}).  We do not include them in our count. Apart from these, the total number of
$\mathbb{Z}_2$ symmetries we find is 2152 (0.2\% of the total) , and the total number of $\mathbb{Z}_3$ symmetries 61664 (6.4\% of the total).   

The $\mathbb{Z}_2$ only occurred in the
class $U(3)\times Sp(2)\times U(1)\times U(1)$ with a massive $B\!-\!L$ boson, as a subgroup of $B\!-\!L$. This is just conventional $R$-parity.
However, the complete set is dominated by models  with a massless $B\!-\!L$ boson, class 2751. Here the standard $\mathbb{Z}_2$ R-parity symmetry is 
already contained in $B\!-\!L$. If we were to exclude all classes with unbroken $B\!-\!L$, there are just $46990$ left. In this set,  about 4.6\% of
the models have a non-trivial  $\mathbb{Z}_2$ discrete symmetry (not including $U(2)_{\bf b}$ duality). But even then, we must conclude that in these models discrete symmetries are a fairly
rare phenomenon, occurring in only a few percent of the cases. The other conclusion is that $\mathbb{Z}_3$ discrete symmetries are about as
common as $\mathbb{Z}_2$ discrete symmetries, at least in this region in the landscape.
In table \ref{Madrid} we list for all 24 classes how many samples there are in the database of \cite{Anastasopoulos:2006da}, and how many
of these have discrete symmetries.

We will now describe each of these cases in a bit more detail. Note that the numbers in table  \ref{Madrid}
specify the number of distinct brane label combinations $({\bf a},{\bf b},{\bf c},{\bf d})$ that yield a given chiral spectrum. Those
spectra are in principle non-chirally distinct (although  in practice there are often huge degeneracies), and also the precise axion couplings may be different. But the differences are small, and hence it suffices 
to present just one example per class. 
In order to make the results reproducible, we specify for each example the tensor product, MIPF and orientifold, and the brane
labels for which they occur. However, it is difficult to present this information in a basis-independent way. Instead we give labels
as used and recognized by the computer program {\tt kac} used to produce these spectra, and which is publicly available. Instructions
for exactly reproducing these spectra can be found on the webpage {\tt http://www.nikhef.nl/$\sim$/t58/Site/String\_Spectra.html}.

\subsubsection{Examples: USUU and UUUU models}

We now turn to several illustrative examples, and their discrete symmetries, which are classified according to the notation in \cite{Ibanez:1991pr}.
To help identify these examples we specify the Hodge numbers of the corresponding Calabi-Yau manifold, by comparing the closed string spectrum
to a type-IIA compactification. In addition, we specify how the $h_{11}$  $N=2$ hyper multiplets split into chiral multiplets and vector multiplets. 
Precise definitions of all these quantities in terms of the partition function are given in the appendix. In each case we indicate which couplings of
phenomenological interest are perturbatively allowed, which ones are forbidden by the discrete symmetry, and which ones are non-perturbatively 
allowed. Couplings in the latter category are not forbidden by any discrete symmetry, and hence can in principle be generated by instantons. However, we are
not claiming that those instantons actually exist in a given model.

\vskip 1.truecm
\leftline{\underline{{\bf Example 1}: $\mathbb{Z}_2$ in $U(3)\times Sp(2) \times U(1) \times U(1)$ with broken $B\!-\!L$ (class 7506)}}
\noindent
An example was found for
tensor product 241446, MIPF 10, Orientifold 2, boundary states $(630,41,1070,631)$. 
The Hodge numbers of the corresponding Calabi-Yau manifold are $h_{21}=28, h_{11}=40$, and in the orientifold $h_{11}^+=35$
(leading to 35 Kahler moduli)  and $h_{11}^-=5$ (5 RR vector bosons).
In this class all Yukawa couplings are perturbatively allowed, as is the $\mu$-term. This is generally true in $USUU$-type Madrid models.
The $\mathbb{Z}_2$  is a a subgroup of the broken $B\!-\!L$, and this
is standard R-parity. 
All dimension-4 baryon and lepton violating couplings are forbidden, including $LH_u$, but the couplings
$QQQL$ and $UUDE$ are non-perturbatively allowed. On the other hand 
Majorana neutrino masses, as well as 
the Weinberg operator $LLH_uH_u$ which can also give rise
to such masses is allowed.  All odd powers of the neutrino superfield are forbidden.
 In this case there are 16 independent axions  (out of a total of 28+1) coupling non-trivially to branes, and
the couplings to the standard model branes is as follows.
\beq
\begin{tabular}{ c c c c c c c c c c c c c c c c c c c c c c c c c c c }
\hbox{{\bf a:}}    &0  &   -3   &0   & 0   & -3  & -3   & -3   & 3   & 0  & -3   & 0   & 3   & 3   & 6   & 3   & 6 \\
\hbox{{\bf c:}}     &0  &   0    &0   & 0   & 0   & 0    & -2   & 0   & 0   & 0   & 0   & 0   & 0   & 0   & 0   & 0 \\
\hbox{{\bf d:}}    &0  & 1     &0   & 0   & 1  & 1   & -1   & -1   & 0  & 1   & 0   & -1   & -1   & -2   & -1   & -2 \\
\end{tabular}
\eeq
\vspace{5pt}

\vskip 1.truecm

\vskip .5cm
 \leftline{\underline{{\bf Example 2}: $\mathbb{Z}_3$ in $U(3)\times Sp(2) \times U(1) \times U(1)$ with broken $B\!-\!L$ (class 7506)}}

\noindent
The $\mathbb{Z}_3$ discrete symmetries in this class do not overlap with the $\mathbb{Z}_2$ symmetries described above.
An example with $\mathbb{Z}_3$ symmetries is
tensor 2101010; MIPF 63; Orientifold 0, boundaries $(192,503,227,237)$ and Hodge numbers
 $h_{21}=7, h_{11}=67$, $h_{11}^+=64$ and $h_{11}^-=3$.
The $\mathbb{Z}_3$ nul vector is (0,1,1), as one can read off from the axion couplings.
\beq
\begin{tabular}{ c c c c c c c c c c c c c c c c c c c c c c c c c c c }
\hbox{{\bf a:}}  & 0 &  0 & -6 &  0 &  3\\
\hbox{{\bf c:}} &   -6 &  6 &  5 & -3 & -4\\
\hbox{{\bf d:}}   & -6 & 6 & 7 &  -3 & -5\\
\end{tabular}
\eeq
\noindent
This symmetry corresponds to  $R_3 L_3^2$ in table 2 of \cite{BerasaluceGonzalez:2011wy}. Note that
the definition of the generator $L$ in this paper differs by a sign from the standard definition of lepton number, which might easily
lead to a confusion between $R_3 L_3$ and $R_3 L_3^2$. This has been taken into account, and
furthermore we have checked explicitly that the discrete symmetries forbid all couplings
of type $UDD$, $QDL$, $LLE$, $LH_u$, $QQQL$, and $UUDE$, confirming that it indeed corresponds to $R_3 L_3^2$.
The $\mu$ term and all Yukawas are perturbatively allowed but neutrino Majorana masses  and the Weinberg operator are forbidden.  
Up to fourth order, there is just one coupling that is perturbatively forbidden but non-perturbatively allowed, and that is the third power of the neutrino superfield.

\vskip 1.truecm

\vskip .5cm
\leftline{\underline{{\bf Example 3}: $\mathbb{Z}_3$ in $U(3)\times Sp(2) \times U(1) \times U(1)$ with unbroken $B\!-\!L$ (class 2751)}}
 \noindent 
 This example occurs for tensor 2101010; MIPF 64 and orientifold 0 and boundaries $(46,5,48,415)$,  Hodge numbers
 $h_{21}=7,  h_{11}=43$, $h_{11}^+=40$ and $h_{11}^-=3$.
  All  $B\!-\!L$ violating couplings are forbidden by the unbroken $B\!-\!L$,
 and the operators $QQQL$ and $UUDE$ are forbidden by the discrete symmetry.  
 All Majorana neutrino mass contributions are also forbidden by $B\!-\!L$.
 The $\mu$-term and all Yukawas are perturbatively allowed. 
 The axion couplings in this example are:
\beq
\begin{tabular}{ c c c c c c c c c c c c c c c c c c c c c c c c c c c }
\hbox{{\bf a:}}  & 9 &  0 & 0 &  0 &  0\\
\hbox{{\bf c:}} &   0 &  0 &  0 & 0 & 0\\
\hbox{{\bf d:}}   & 3 & 0 & 0 &  0 &  0
\end{tabular}
\eeq
\\
This obviously has an $L_3$ discrete symmetry: lepton number  can only be violated in units of three.
Since $B\!-\!L$ is conserved the same is then automatically true for baryon number as well. Note that
$Y$ conservation forces the {\bf a}-brane couplings to be three times those of the ${\bf d}$. The resulting
factor $9$ incorporates both $SU(3)$ triality and the $B_3$ discrete symmetry.

\vskip 1.truecm

\vskip .5cm
\leftline{\underline{{\bf Example 4}: $\mathbb{Z}_3$ in $U(3)\times U(2) \times U(1) \times U(1)$ with unbroken $B\!-\!L$ (class 1352)}}
An example of this kind occurs for tensor 2101010; MIPF 59; Orientifold 0, boundaries  $(932,650,881,1302)$.
The Hodge numbers are $h_{21}=19,h_{31}=59$, $h_{11}^+=29$ and $h_{11}^-=2$.
The axion couplings are:\\
\beq
\begin{tabular}{ c c c c c c c c c c c c c c c c c c c c c c c c c c c }
\hbox{{\bf a:}} &    0 & 0 & 0 & 9 & 0 & 0 & 0 & 9 & 0 & 0 & 9 & 0 & 0 & 0 & 9 & 9 & 9\\
\hbox{{\bf b:}} & 2 & 2 & 2 & 4 & 4 & 2 & 0 & 0 & 4 & 0 & 2 & 0 & 2 & 2 & 0 & 0 & 0\\
\hbox{{\bf c:}} &  0 & 0 & 0 & 0 & 0 & 0 & 0 & 0 & 0 & 0 & 0 & 0 & 0 & 0 & 0 & 0 & 0\\
\hbox{{\bf d:}} &   0 & 0 & 0 & 3 & 0 & 0 & 0 & 3 & 0 & 0 & 3 & 0 & 0 & 0 & 3 & 3 & 3
\end{tabular}
\eeq
\vspace{5pt}\\
from which we can read off that there is indeed a massless $Y$ and $B-L$, and that
furthermore there is a 
$\mathbb{Z}_2$  null vector      $(0,    1,    0,    0)$ and a 
$\mathbb{Z}_3$  null vector      $(0,    0,    0,    1)$. The former just imposes $SU(2)$ duality,
and the second corresponds to an $L_3$ discrete symmetry, as in example 3. 
In this class of models, two of the three quark masses (for up as well as down quarks)
must be generated non-perturbatively, but the discrete symmetries do not forbid that.
All lepton Yukawas are perturbatively allowed.
We are assuming here, as in the USUU examples, that the Higgs comes from the
non-chiral spectrum, from bi-fundamentals between the ${\bf b}$ and ${\bf c}$ branes.
The $B\!-\!L$ forbids the usual dimension-4 terms as well as Majorana neutrino masses, while $QQQL$ and $UUDE$ are 
forbidden by the $\mathbb{Z}_3$ symmetry. A $\mu$-term can in principle be generated
non-perturbatively, and is not forbidden by the $\mathbb{Z}_3$.

\vskip 1.truecm

 \leftline{\underline{{\bf Example 5}: $\mathbb{Z}_3$ in $U(3)\times U(2) \times U(1) \times U(1)$ with unbroken $B\!-\!L$ (class 7976)}}
 
 This example occurs for tensor 242222; MIPF 16; orientifold 2 and boundaries $(343,6,610,436)$, Hodge numbers
 $h_{21}=21, h_{11}=69$, $h_{11}^+=52$ and $h_{11}^-=17$.
 The axion couplings are
\beq
\begin{tabular}{ c c c c c c c c c c c c c c c c c c c c c c c c c c c }
\hbox{{\bf a:}}  & 0 &  0 & 18 &  0 &  0 &  0 &  0 &  9 &  9  & -18 &  0 &  0 & -9 &  9 &  9\\
\hbox{{\bf b:}}   & 0 &  0 &  0 &  0 &  0 &  0 &  0 &  0 &  0 &  0 &  0 &  0 &  0 & -2 &  0\\
\hbox{{\bf c:}}   & 0 &  0 &  0 &  0 &  0 &  0 &  0 &  0 &  0 &  0 &  0 &  0 &  0 &  0 &  0\\
\hbox{{\bf d:}}   & 0 &  0 & -6 &  0 &  0 &  0 &  0 & -3 & -3 &  6 &  0 &  0 &  3 & -3 & -3\\
\end{tabular}
\eeq
 The rest of the discussion is similar for example 4. There is a massless $B\!-\!L$ forbidding
 the usual couplings, as above.
 However, in this case the Higgs pair
 comes out automatically within the chiral spectrum.  Perturbative Yukawa couplings 
 appear for 
 two of the up and down-quarks and one of the leptons, and the missing ones are
 in principle allowed non-perturbatively.
 Also a $\mu$-term may be generated non-perturbatively. 
 However all $QQQL$ and $UUDE$ terms are forbidden by the discrete symmetries.

 \vskip 1.truecm
 \leftline{\underline{{\bf Example 6}: $\mathbb{Z}_3$ in $U(3)\times U(2) \times U(1) \times U(1)$ with unbroken $B\!-\!L$ (class 14792)}}
The next example also has a $\mathbb{Z}_3$ symmetry, but it is a bit different from the foregoing two. 
It occurs for
tensor 242222; MIPF 16; orientifold 2, boundaries $(284,343,700,335)$,  Hodge numbers
 $h_{21}=21, h_{11}=69$ , $h_{11}^+=52$ and $h_{17}^-=2$ (as in the previous example).
\nobreak
\beq
\begin{tabular}{ c c c c c c c c c c c c c c c c c c c c c c c c c c c c c c c c c c c c }
\hbox{{\bf a:}} &   -3 &  -3 &  27  &  3  &  9 &  -6 &  -6  &  18 &  21 & -33  &  0  &  6 &  -9  &  0  &  3\\
\hbox{{\bf b:}} &    0 &  0  &  12   &  0  &  0  &  0  &  0  &  6    &  6   & -12  &  0  &  0 &  -6  &  6  &  6\\
\hbox{{\bf c:}} &    0 &  0  &  0  &  0  &  0  &  0  &  0  &  0  &  0  &  0  &  0  &  0  &  0  &  0  &  0\\
\hbox{{\bf d:}} &   -1 &  -1  &  9  &  1  &  3 &  -2 &  -2  &  6  &  7 & -11  &  0  &  2 &  -3  &  0  &  1
\end{tabular}
\eeq 
In this example, the {\bf b}-brane couplings have a $\mathbb{Z}_6$ symmetry. The $\mathbb{Z}_2$
subgroup is the usual uninteresting  $SU(2)$ duality. However, the $\mathbb{Z}_6$ symmetry forbids
any operator built out of two or four $U(2)$ doublets with the same $U(1)$ charge, such as $QQQL$. On the other hand, it does not
forbid the operator $UUDE$. The dimension-4 operators are forbidden by the unbroken $B\!-\!L$. This discrete symmetry is not
explicitly mentioned in \cite{BerasaluceGonzalez:2011wy} because these authors allowed at most one $U(2)$-chiral Higgs pair,
and this model has six. This discrete $\mathbb{Z}_6$ symmetry is anomaly free as long as the number of families is a multiple
of three, and the number of Higgs doublets a multiple of six, which is the case here (if it were not the case a $\mathbb{Z}_6$ 
symmetry could not
have appeared). We assume that one of these six Higgs candidates plays the r\^ole of the Higgs boson. Then all quark and
lepton Yukawa couplings are perturbatively allowed. A $\mu$-term however cannot be generated even non-perturbatively because
of the discrete symmetry.

\vskip 1.truecm
 \leftline{\underline{{\bf Example 7}: $\mathbb{Z}_3$ in $U(3)\times U(2) \times U(1) \times U(1)$ (exotic)  (class 7488)}}
This is an example of discrete symmetries occurring for a non-Madrid model. 
It was found for tensor 441010; MIPF 21; Orientifold 1, Hodge numbers $h_{21}=28, h_{11}=10$,  $h_{11}^+=10$ and $h_{11}^-=0$
\nobreak
\beq
\begin{tabular}{ c c c c c c c c c c c c c c c c c c c c c c c c c c c c c c c c c c c c }
\hbox{{\bf a:}} & -3 &  -6 &  -3 & 3  & 0 & 0\\
\hbox{{\bf b:}} &  -4 &  -2 &  6 & 0 &  -4  & 2\\
\hbox{{\bf c:}} &  1 &  -1 &  -1 &   1 &  -1 &  -1\\
\hbox{{\bf d:}} & 2 &  1 &  0 &  0 &  -1 &  -1\\
\end{tabular}
\eeq
The only massless $U(1)$ is $Y$. Furthermore, we find the usual $SU(2)$-duality
$\mathbb{Z}_2$, but also a 
$\mathbb{Z}_3$ null vector is $(0,1,0,2)$. Note that the {\bf d}-brane cannot be associated with
lepton number, since this is not a Madrid model. Furthermore there is no family universality
for the up and down quarks. This has the consequence that there is no universal rule for 
the dimension 4 couplings: some of the $UDD$ couplings are non-perturbatively allowed, and
some are forbidden by discrete symmetries; some of the $QDL$ couplings are perturbatively allowed,
others are non-perturbatively allowed, and some are forbidden by the discrete $\mathbb{Z}_3$ symmetry.
All $LLE$ coupling are non-perturbatively allowed. Some of the $QQQL$ and $UUDE$ couplings are non-perturbatively
allowed, and some are forbidden by the $\mathbb{Z}_3$ symmetry. We cannot discuss Higgs couplings in general, because
none of the particles in the Chan-Paton chiral spectra can play the r\^ole of the Higgs. The three particles in the {\bf bc}
intersection must be interpreted as lepton doublets. Clearly the Higgs must come from the non-chiral spectrum, and could come
from any strings ending on the ${\bf b}$ brane en with the other end on the ${\bf c}$  or ${\bf d}$ brane. Since lepton number is
undefined, there is no obvious way of deciding this.
The precise phenomenological fate of this type of model is hard to assess, because of the family-dependent presence of
baryon and lepton number violating couplings, and the fact that there is no obvious Higgs candidate.

\subsection{GUT spectra}

Orientifold spectra with an $SU(5)$ GUT  spectrum have been studied in many papers
\cite{cv1,susy,non-susy,bere,Kiritsis:2009sf,Anastasopoulos:2010hu}. In \cite{BerasaluceGonzalez:2011wy}
the possible presence of discrete symmetries in models was discussed. There are two important issues here, one in
which discrete symmetries would be catastrophic, and another where they would be beneficial. The first concerns the up-quark
Yukawa couplings\footnote{In flipped $SU(5)$ models these remarks apply to the down-quark Yukawa couplings} which in these models 
are forbidden perturbatively, but may be generated by instantons \cite{Blumenhagen:2007zk,Ibanez:2008my}.
A discrete symmetry might forbid the existence of these instantons. On the other hand, discrete symmetries to forbid
the usual dimension four $B$ and/or $L$ violating couplings would be  very welcome. 
For $SU(5)$ GUT models baryon number
violating MSSM couplings are {\it a priori} even more threatening , because the same instantons that generate the Yukawa
couplings also tend to generate the $QQQL$ term \cite{Kiritsis:2009sf}. Here discrete symmetries are less obviously useful, because
they would tend to forbid the up-quark Yukawas as well. However, in \cite{Anastasopoulos:2010hu} it was pointed out that specially
chose brane realizations can help solving this problem.

Here we examine the presence of discrete symmetries for certain Gepner
model GUT realizations, namely all the two-stack models where the second stack is either $U(1)$ or $O(1)$, and
a subset of the three-stack models.

\subsubsection{Two-stack models}

The simplest realizations of $SU(5)$ GUT consist of one stack of five branes producing a $U(5)$ Chan-Paton group, and
a second stack which does not couple to any of the standard model gauge interactions. The chiral matter consists of three 
$({\bf 10})$'s of $SU(5)$, plus three bi-fundamentals from open strings stretched between the two stacks that are in the $({\bf 5}^*)$
of $SU(5)$. In addition there may be a number of $({\bf 5})+({\bf 5}^*)$ pairs that play the r\^ole of Higgs bosons. 
In the database of \cite{Anastasopoulos:2006da} there are five distinct types of such models with an $U(1)$ or $O(1)$ second stack. These were discussed
in some detail in \cite{Kiritsis:2009sf}. 
There also exist
two-stack models where the second stack is $U(3)$ or $O(3)$, with the multiplicity of the $({\bf 5}^*)$ originating from the extra
brane group, but this leads to additional complications, and we will not consider them here. The chiral spectrum of the five classes is shown in table
(\ref{SU5}).

We have examined this entire class for the presence of discrete symmetries, and
found only a few examples for one of the $U(5)\times U(1)$ classes, namely nr. 345 (In \cite{Kiritsis:2009sf} the models were numbered
by frequency, and this class was
referred to as nr. 2753, 
with a total of 1136 samples in the database). The discrete symmetry is a $\mathbb{Z}_2$ embedded in the two $U(1)$ factors. 
However, the chiral matter is in antisymmetric tensors of $U(5)$,  bi-fundamentals and symmetric tensors of $U(1)$ which are
all uncharged with respect to this $\mathbb{Z}_2$. Hence there is no chance of forbidding any couplings. This implies on the one
hand that there is no obstacle to generating the perturbatively forbidden up-quark Yukawa couplings, but on the other hand
that there is no chance of forbidding dangerous $B$ and/or $L$ violating operators. Examples of instantons generating these
Yukawas were indeed found in \cite{Kiritsis:2009sf}, but only in six case tadpole canceling hidden sectors could be added, and
only by allowing chiral observable-hidden matter (which becomes vector-like when reduced to the standard model).

\subsubsection{Three-stack models}

If we allow two additional stacks instead of one the number of possibilities becomes much larger. There are 257 classes with
a Chan-Paton group $U(5) \times U(1) \times U(1)$ and 168 with a group $U(5) \times U(1) \times O(1)$. If we allow higher multiplicities for
the extra branes there are even more possibilities. Here we will only consider the subclass studied in \cite{Anastasopoulos:2010hu}, in which
the up-quark Yukawa couplings can be generated by instantons without generating $QQQL$ couplings. This class consists of seven chiral types,
listed in table \ref{SU5}. All models in the class considered here  have natural Higgs candidates in the chiral spectrum coming from 
${\bf a}^*{\bf b}^*$ and ${\bf a}{\bf c}$ bi-fundamentals. There are always three Higgs pairs.
Here we chose to assign the ${\bf a}^*{\bf b}$ bi-fundamentals to the $({\bf 5}^*)$ of $SU(5)$ containing the lepton doublet and the anti-down quarks.
Note that there are always three mirror pairs of particles with the quantum numbers of down quarks which form an $SU(5)$ multiplet together with
the three Higgs pairs. This is the usual doublet-triplet splitting problem. In analyzing couplings, we have to make sure to take the down quarks from
the ${\bf a}^*{\bf b}$ bi-fundamentals, and assume that the others pair off into massive particles. Matter from the ${\bf bc}$ sector is neutral, and
could play the r\^ole of neutrinos, although there are either six or eight such states. In addition, in model 4325 there is a symmetric tensor, which
is another neutrino candidate. Note that the anti--symmetric tensors on the ${\bf b}$ and ${\bf c}$ branes listed in table \ref{SU5} correspond
to string sectors without massless states. They are merely listed here because there do exist massive states in these sectors, and because they
contribute to anomaly cancellation.

\subsubsection{Examples: GUT models}

Here we present for each distinct class of discrete symmetries one example in some detail. The model numbers (``class") refer to table  \ref{SU5}.
Note that the discrete symmetries in the only two-stack example in table \ref{SU5}, model 345, do not forbid anything, as explained above. So we
do not present that example in more detail. 
\vskip .5cm
 \leftline{\underline{{\bf Example 8}: $\mathbb{Z}_2$ in $U(5)\times U(1) \times O(1)$  (class 57)}}

This was found for tensor 1102222; MIPF 18; orientifold 0 with boundaries numbers $(566,566,1308,990)$.  The Hodge numbers are $h_{21}=13, h_{11}=109$,  $h_{11}^+=98$ and $h_{11}^-=11$.
Note that the  first two boundary labels are  here for the standard model $U(3)$ and $U(2)$ stacks. The fact that they are identical implies that we have $U(5)$. In the following examples we will always
combine these two groups to $U(5)$ when writing axion couplings. Then {\bf a} denotes the $U(5)$ stack, {\bf b} the first $U(1)$, and {\bf c} the second $U(1)$ (if any).
The axion couplings are
\nobreak
\beq
\begin{tabular}{ c c c c c c c c c c c c c c c c c c c c c c c c c c c c c c c c c c c c }
\hbox{{\bf a:}} &  5 &  0 & -10 &  0 &  5 & -10 & -5 &  5 &  -20 &  0 \\
\hbox{{\bf b:}}  &  0 & -2 & -2 &  0 &  0 &  0 &  0 &  0 &  0 &  0\\
\end{tabular}
\eeq
There is a surviving $\mathbb{Z}_2$ symmetry associated with the $U(1)$ brane.
This symmetry forbids the $LH_u$ term, the $\mu$-term, as well as all $QQQL$ and $UUDE$
terms. Down quark and lepton Yukawa couplings are perturbatively allowed, and up-quark
Yukawas are non-perturbatively allowed. But on the other hand, $UDD$, $QDL$ and $LLE$ terms
are non-perturbatively allowed as well, although they are generated by different instantons then
the Yukawa couplings, and hence could have a different strength. Since the $\mu$ term is perturbatively
forbidden, it follows from $SU(5)$ symmetry that also a mass term for the color triplet partners of the Higgses
is forbidden. So here the discrete symmetry has a negative effect. In the down quark sector we get $6 (d^*) + 3 (d)$, 
and all options for pairing off the vector-like $d$-quarks are forbidden by the discrete symmetry.

\vskip .5cm
 \leftline{\underline{{\bf Example 9}: $\mathbb{Z}_3$ in $U(5)\times U(1) \times U(1)$  (class 4004)}}

This was found for tensor 1102222; MIPF 28, orientifold 0 and boundaries numbers $(816,816,1309,917)$.  
The Hodge numbers are $h_{21}=13, h_{11}=37$,  $h_{11}^+=34$ and $h_{11}^-=3$. The axion couplings are
\nobreak
\beq
\begin{tabular}{ c c c c c c c c c c c c c c c c c c c c c c c c c c c c c c c c c c c c }
\hbox{{\bf a:}}   & 0 &  5 & -10 &  -5 &  0 &  5 &  5 & -10 &  0 & -5\\
\hbox{{\bf b:}}   &  0 &  0 &  0 &  0 &  0 &  0 &  0 &  0 &  0 &  0\\
\hbox{{\bf c:}}   &  0 &  1 &  1 & -1 &  0 &  1 &  1 & -2 &  0 & -1\\
\end{tabular}
\eeq
Note that the vector boson coupling to $Y_b$ remains massless. This forbids many superpotential terms. 
There is a surviving $\mathbb{Z}_3$ symmetry 
embedded in the combination of the $U(1)$-generator of the ${\bf a}$ stack and the
one of the  ${\bf c}$ stack. This symmetry forbids masses for the neutrino candidates from the {\bf b}-{\bf c} intersection.
Majorana masses for these six states are already forbidden by the masses $Y_{\bf b}$, but the discrete symmetry also
prevents them from pairing off into a Dirac mass term. Nothing else of any interest is forbidden by the discrete symmetry.
Yukawa couplings are perturbatively or non-perturbatively allowed, as in the previous case.

\vskip .5cm
 \leftline{\underline{{\bf Example 10}: $\mathbb{Z}_2$ in $U(5)\times U(1) \times U(1)$  (class 4316)}}

It was found for tensor 1102222; MIPF 18, orientifold 0, and boundary numbers $(523,523,1307,566)$, Hodge numbers $h_{21}=13, h_{11}=109$,  $h_{11}^+=98$ and $h_{11}^-=11$. The axion couplings are:
\nobreak
\beq
\begin{tabular}{ c c c c c c c c c c c c c c c c c c c c c c c c c c c c c c c c c c c c }
\hbox{{\bf a:}} &   5 &  0 & -10 &  0 &  5 & -10 & -5 &  5  & -20 &  0 \\
\hbox{{\bf b:}} &   0 &  2 &  2 &  0 &  0 &  0 &  0 &  0 &  0 &  0\\
\hbox{{\bf c:}} &   1 &  0 & -2 &  0 &  1 & -2 & -1 &  1 & -4 &  0\\
\end{tabular}
\eeq
This case is similar to the foregoing one, except that $Y_{\bf a} - 5 Y_{\bf c}$ rather than $Y_{\rm b}$ remains massless.
There is a surviving $\mathbb{Z}_2$ symmetry 
in $Y_{\rm b}$, but no couplings are forbidden by it that are not already forbidden by the extra $U(1)$. Interestingly, {\it all} fourth order superpotential terms 
are absent in this class of models, as well as all first and second order terms. The absence of second order terms implies in particular that mass terms 
for the vector-like down quarks from the Higgs multiplets are forbidden. These features are a property of the entire class, irrespective of discrete symmetries.
The problem is that there are $(5^*)$'s from {\bf a}$^*${\bf b} and  {\bf a}$*${\bf b}$^*$ and a $(5)$ from {\bf a}{\bf c}. The first two have charge $-1$ w.r.t.
$Y_{\bf a} - 5 Y_{\bf c}$, but the latter has charge $-4$. This forbids any Dirac mass term.
Therefore to make this kind of spectrum viable we first have to break the additional $U(1)$ symmetry, either by a Higgs mechanism, or by axion mixing
directly in string theory. 
There are indeed example in the database where the latter occurs. This is class 4324, which will be discussed below.

\vskip .5cm
 \leftline{\underline{{\bf Example 11}: $\mathbb{Z}_3$ in $U(5)\times U(1) \times U(1)$  (class 4316)}}

This example was found for tensor 441010; MIPF 71; orientifold 0 and boundary states $(485,485,525,581)$, Hodge numbers $h_{21}=20, h_{11}=14$,  $h_{11}^+=14$ and $h_{11}^-=0$. The axion couplings are
\nobreak
\beq
\begin{tabular}{ c c c c c c c c c c c c c c c c c c c c c c c c c c c c c c c c c c c c }
\hbox{{\bf a:}} & -5 & -10 & -5 & -10 &  5 &  0 &  5 & 20 & -10 & -10 \\
\hbox{{\bf b:}} &   0 &  0 & -3 &  0 &  0 &  0 &  0 &  0 &  0 &  0\\
\hbox{{\bf c:}} & -1 & -2 & -1 & -2 &  1 &  0 &  1 &  4 & -2 & -2\\
\end{tabular}
\eeq
This is  similar to the previous example, except that the discrete
symmetry that is embedded in $U(1)_{\rm b}$ is $\mathbb{Z}_3$. As before, the superpotential only
contains terms of order three, or five and higher, as a consequence of the extra $U(1)$. Hence we have
the same problem with lifting the vector-like down quark pair. 
The discrete symmetry forbids the terms $UDD$, $QDL$
and $LLE$, provided that for $D$ we use the field that is {\it not} in the Higgs multiplet. There are even perturbatively
allowed baryon number violating couplings involving the vector-like down quark pair.
Up-quark Yukawa
couplings are non-perturbatively allowed, down quark Yukawas are perturbatively allowed, as are charged lepton
Yukawas, and neutrino Yukawa couplings are forbidden by the discrete symmetry.

\vskip .5cm
 \leftline{\underline{{\bf Example 12}: $\mathbb{Z}_2$ in $U(5)\times U(1) \times U(1)$  (class 4324)}}

This was found for tensor 1102222; MIPF 27; Orientifold 0 and boundary states  
$( 365, 365,  1393, 572)$ with Hodge numbers $h_{21}=12, h_{11}=96$,  $h_{11}^+=90$ and $h_{11}^-=6$.
This class is similar to 4316 discussed above, except that there are no extra $U(1)$ gauge bosons, and
consequently many more couplings are allowed. In particular this includes the vector-like 
down-quark pair. Since in model 4316 this mass was forbidden by the extra $U(1)$, it follows that in class 4324
it may in principle be generated by instantons. However, that cannot happen in the eight cases with extra discrete symmetries
we are discussing here. In this particular example
the axion couplings are: 
\nobreak
\beq
\begin{tabular}{ c c c c c c c c c c c c c c c c c c c c c c c c c c c c c c c c c c c c }
\hbox{{\bf a:}} &  0 &  5 & -5 & -10 & -5 &  0 & -10 &  0 &  5\\
\hbox{{\bf b:}} &  2 &  0 &  0 &  0 &  0 &  0 &  0 &  0 &  0\\
\hbox{{\bf c:}} &   0 &  0 & -1 & -2 & -1 &  0 & -2 &  0 &  1\\
\end{tabular}
\eeq
{}From which we read off a discrete
symmetry $\mathbb{Z}_2$ associated with brane {\bf b}. This symmetry forbids
 the mass terms needed to lift the vector-like down-quark. It also forbids a $\mu$-term, and all $QQQL$ and
 $UUDE$ terms. The up-quark Yukawas are non-perturbatively allowed, and the down quark and 
 charged lepton Yukawas are perturbatively allowed. Note that all three-stack models we consider here
 satisfy a criterium discussed in \cite{Anastasopoulos:2010hu}, namely that up-quark Yukawas can be
 generated without automatically generating $QQQL$ and $UUDE$ term of similar strength. This is generically
 a problem in $U(5)$ orientifold models \cite{Kiritsis:2009sf}. In the three-stack classes discussed here this problem
 can be solved because the up-quark Yukawas and the $QQQL$ and $UUDE$ terms are generated by instantons
 with different charges. Hence it is at least possible in principle that they contribute with different strengths.
 Here we see an even better solution to this particular problem. Precisely because these terms violate a different
 set of charges, it is possible for discrete symmetries to forbid one and not the other. That is exactly what is happening 
 here. So we see here an example where discrete symmetries are playing a very useful r\^ole, but this is overshadowed by
at least  two serious problems. The first is that there are {\it no} discrete symmetries forbidding the $UDD$, $QDL$ and $LLE$
terms. These are perturbatively forbidden, but may be generated by instantons. The second is the forbidden vector-like down-quark 
mass ({\it i.e.} the down quark triplets in the Higgs multiplets). The brane charge violation of the latter terms is precisely the sum of the charge violations of the up-quark Yukawas and the
 $QQQL$ or $UUDE$ terms. Hence any discrete symmetries that  forbid  the latter but not the former will automatically forbid the lifting of the down quark mirror pair.
 This is just a manifestation of the  doublet-triplet splitting problem in $SU(5)$ models.

\vskip .5cm
 \leftline{\underline{{\bf Example 13}: $\mathbb{Z}_2$ in $U(5)\times U(1) \times U(1)$  (class 4325)}}

This was found for tensor 1102222; MIPF 27; Orientifold 0 and boundary states  $( 365, 365,  1506,   818)$,  with Hodge numbers $h_{21}=12, h_{11}=96$,  $h_{11}^+=90$ and $h_{11}^-=6$.
This class is very similar to 4324. It has some additional neutral chiral matter, but as in class 4324 there is no additional massless $U(1)$.
The axion couplings are
\nobreak
\beq
\begin{tabular}{ c c c c c c c c c c c c c c c c c c c c c c c c c c c c c c c c c c c c }
\hbox{{\bf a:}} &  0 &  5 & -5 & -10 & -5 &  0 & -10 &  0 &  5\\
\hbox{{\bf b:}} &  1 &  0 &  0 &  0 &  0 &  0 &  0 &  0 &  0\\
\hbox{{\bf c:}} &   -1 &  1 & -3 & -2 & -1 &  0 & -2 &  0 &  1\\
%
\end{tabular}
\eeq
The discrete
symmetry is $\mathbb{Z}_2$. The corresponding null vector is $(1,1,1)$. This implies that all matter is uncharged with respect to 
it, because all matter is either in rank two tensors or bi-fundamentals. Hence no couplings are affected by this symmetry. 
One point worth noting is that if there are hidden sectors, any observable-hidden matter is necessarily odd under the $\mathbb{Z}_2$.
Hence this symmetry is like an exotic ({\it i.e} observable-hidden) matter parity\rlap.\footnote{The same remark applies to the
$\mathbb{Z}_2$ symmetry we found in the two-stack models in class 345, 
but which was not presented in detail.} 
All exotic matter can only be created in pairs, and there
will be  a lightest exotic state that cannot decay into standard model particles, and hence, if neutral,  could be a dark matter candidate. 
Note that in the $U(5)$ class observable-hidden matter has integral electric charge (whereas in Madrid type models they have half-integer charge),
and hence this conserved exotic matter parity is not a trivial consequence of charge conservation, nor is it in disagreement with the fact
that no fractional electric charge has ever been observed. This mechanism could in principle work equally well in non-supersymmetric models
(provided examples can be found) and hence this provides an alternative to the r\^ole of R-parity in solving the dark matter problem. Furthermore
the general category  to which this model belongs (the ``$x=0$" category of \cite{Anastasopoulos:2006da}) includes plenty of examples where instead of a 
$U(5)$ stack there are separate $U(3)$ and $U(2)$  stacks, so that there is no $SU(5)$ relation among the couplings (which would be a problem
without low energy supersymmetry). However,  unlike R-parity, the existence of this exotic matter parity is lacking  a convincing
motivation.

\subsection{Tadpole Cancellation}

All models we have considered so far are brane configurations, which in most cases have uncancelled tadpoles. 
For all cases where non-trivial discrete symmetries were found ({\it i.e.} those listed in the last columns of table \ref{Madrid} and \ref{SU5}) we have
attempted to find hidden sectors that cancel all tadpoles, allowing at most four additional stacks. Furthermore we have allowed massless matter in the
observable-hidden sector, provided that it is non-chiral with respect to the full Chan-Paton gauge group, so that in principle it can acquire a mass 
by moving into moduli space, without breaking any gauge symmetries. These are essentially the same criteria used in   \cite{Dijkstra:2004cc} and 
 \cite{Anastasopoulos:2006da}, except that in those searches the number of additional branes was only limited by practical considerations\rlap.\footnote{In all these cases cancellation of 
K-theory charges was checked  \cite{GatoRivera:2005qd} after the results \cite{Dijkstra:2004cc} were published, which is the reason why
there are a few minor discrepancies between the numbers quoted here and those listed in  \cite{Dijkstra:2004cc}. All brane configurations in the
database of \cite{Anastasopoulos:2006da} satisfy all K-theory constraints that can be obtained using RCFT probe branes \cite{Uranga:2000xp}. }
Among the non-Madrid models, only classes 14062 and 7488 contain at least one model allowing a tadpole-cancelling hidden sector. 
 
 Based on previous experiences with solving tadpole conditions in RCFT orientifolds, we expected only a relatively small success rate, around one 
 percent or even less. Surprisingly however, the overall success rate for the models in table \ref{Madrid} was much higher, around $65\%$. We found a 
 total of 41456 cases with tadpole-cancelling hidden sectors, for a total of 63728 configurations with discrete symmetries. On the other hand, in the
 class of $SU(5)$ models, we did not find a single case with tadpole cancellation and discrete symmetries.

 One might be tempted to conclude, although discrete symmetries are rare within the set of standard model configurations, they are more common 
 in the physically relevant class of fully consistent open string models. There is indeed a reason why that could be true. Both tadpole cancellation and
 discrete symmetry conditions are more easily satisfied if there are fewer Ishibashi states reps. axions, which correlates with smaller values of $h_{12}$.
 Hence it is quite easily imaginable that after imposing the requirement of having discrete symmetries, we are left with precisely those cases where
 the tadpole conditions are more easy to satisfy as well.
 
 On the other hand, it turns out that especially in the set of 59808 spectra there are huge degeneracies. As already observed in \cite{Dijkstra:2004cc}, in many cases
 different boundary state combinations in the same orientifold yield the same spectrum, also for non-chiral states. The origin of these degeneracies is not
 understood. In the case of
 MIPFs and orientifolds, permutation degeneracies were removed, and this is usually enough to be left with a set of truly distinct ones. We did not attempt to
 remove permutation degeneracies for boundary states, but in any case that would not be enough to understand the remaining apparent degeneracies. Furthermore
 we do no know if the apparent degeneracies are genuine. For example, it is possible that there are slight differences in the set of available instanton branes and
 their zero-modes, which altogether provides a huge number of parameters that could be compared. But in any case, it is an empirical fact that the number of
 tadpole free spectra for a given orientifold can exceed the number of {\it distinct} tadpole free spectra by a factor of one hundred or more. Furthermore, even
 non-degenerate spectra often have only minor differences in their vector-like states, suggesting that they are nearby points in the open string moduli space.
 In particular, if one of them satisfies the tadpole conditions for some hidden sector, usually its close relatives have the same property. Even if these degeneracies were
 fully understood, it is not obvious how to take them into account properly in comparing frequencies of certain features of interest. Therefore statements
 about this should only be taken as a rough indication.
 
 It turns out that of the 41136 tadpole-free spectra, 31016 come from just one orientifold, and another 9792 from another one. This strongly suggests that two
 sets of near-degenerate cases dominate the entire sample.  We are unable to decide whether this is accidental or whether this should be seen as 
 support for the idea that the presence of discrete symmetries enhances the chance of satisfying the tadpole conditions. 
 Furthermore, we did not attempt to solve the tadpole conditions in those cases where we did not find
 discrete symmetries. 
 For these reasons, we cannot give a reliable estimate of the likelihood of having both tadpole cancellation and discrete symmetries.

\section{Conclusions}

In this paper we have studied the appearance of discrete ${\mathbb{Z}_N}$ gauge symmetries within a large class of
RCFT Type II  4d orientifolds with a MSSM-like spectrum.  Although interesting for its own sake, our study
is motivated by the fact that such discrete symmetries like R-parity or other ${\mathbb{Z}_N}$ generalizations
are necessary to avoid large baryon and/or  lepton number
violation in the MSSM.  Their presence also dictates the possible signatures of low energy SUSY at the LHC.
Thus e.g. if the  lightest SUSY particle is not stable and its decay violates lepton number , the search for squarks
and gluinos at LHC through R-parity violation channels  leads to much weaker mass limits compared to 
those preserving R-parity.

So a first question is whether indeed such discrete gauge symmetries arise naturally in string 
compactifications. We have done a systematic search for such symmetries in one of the largest 
sets available of 4d compactifications with three generations and a MSSM-like structure. 
These are RCFT Type II orientifold models with modular invariant partition functions
based on Gepner models. 

One of our most important results is that we have explicitly constructed, for all Gepner orientifolds that can be obtained
with simple current MIPFs and all orientifold projections of \cite{Fuchs:2000cm}, an integral basis
for couplings of axions with  all complex branes. This allows us to investigate, under a plausible
assumption, whether the $U(1)$'s in a given model are broken completely, or broken to a discrete
subgroup. This analysis can easily be performed for any Gepner orientifold, and all discrete 
subgroups
of the full set of $U(1)$'s can be determined in this manner. The integral basis itself may give insight
in the geometrical structure underlying to these RCFT models, but we will not explore  this issue here.

In this large class of MSSM-like models we did indeed find cases with the appropriate discrete gauge symmetries. 
The only non-trivial ones were $\mathbb{Z}_2$ (which turned out to be standard MSSM R-parity) and some other
 $\mathbb{Z}_3$ symmetries, not including baryon triality. In models with an additional $U(1)_{B-L}$ there appear symmetries forbidding dimension 5 baryon number violating operators.
The finding of these discrete symmetries in string compactifications is, in one hand, good news for theories
of low energy SUSY like the MSSM. It shows that symmetries like R-parity, which are imposed in the MSSM
in a totally ad-hoc fashion can find a more fundamental origin as discrete remnants $U(1)$ symmetries.

The examples discussed in detail in the previous section display a wide range of positive and negative effects 
discrete symmetries may have. Basically any effect that was foreseen does indeed occur, with good or bad
consequences for dimension four or five B and/or L violating couplings, Yukawa couplings,
neutrino masses or the $\mu$-term. In addition one example showed an unexpected feature, namely a discrete
$\mathbb{Z}_2$ symmetry under which fields with both visible and hidden sector quantum numbers are odd.
Neutral fields of this kind could provide for new candidates for dark matter.

On the other hand the presence of such symmetries does not seem generic, at least within the studied
class of compactifications, they only appear at a few percent level for MSSM-like configurations. 
However, it is worthwhile to point out his relative frequency is sizably above what would be obtained from complete randomness, i.e. by considering the axion coupling coefficients as random variables, and computing the probability of them having a given common factor (for instance, for say 10 active axions, the random probability of a $\mathbb{Z}_3$ can be estimated as $(1/3)^{10}\sim 10^{-5}$). A final remark is that the fraction of RCFT models with discrete symmetries could be enhanced after imposing additional theoretical
constraints; for instance we indeed find some indications
that the percentage goes up if one considers tadpole-free models.  It would certainly be worth studying 
how often the required discrete symmetries appear in other large classes of string compactifications.

\vspace{1.5cm}

\centerline{\bf \large Acknowledgments}

\bigskip

We thank  F. Marchesano  for useful discussions.  
This work has been partially supported by the grants FPA 2009-09017, FPA 2009-07908 and Consolider-CPAN (CSD2007-00042)  from the Spanish Ministry of Economy and Innovation, HEPHACOS-S2009/ESP1473 from the C.A. de Madrid and the contract ``UNILHC" PITN-GA-2009-237920 of the European Commission.

\newpage

\appendix

\section{Hodge numbers}

In order to help identifying the orientifolds used in the examples,
in comparison with geometric data, we specify for each model the Hodge numbers of the corresponding type-II theory as well
as their orientifold projection.  In this appendix we explain precisely how these number are computed from the
RCFT data.

The standard way of computing the 
spectrum of a Gepner model is by diagonally combining all minimal model and NSR characters. This leads to
an $N=2$ spectrum with certain numbers of hyper multiplets and vector multiplets. These spectra can often be identified with
those of ten-dimensional strings compactified on Calabi-Yau manifolds, a statement that can be made precise using the
Landau-Ginzburg correspondence \cite{Greene:1988ut,Vafa:1988uu,Lerche:1989uy,Vafa:1989xc}. 

Characters of the chiral algebra (including space-time and word-sheet
supersymmetry extensions) of Gepner models that have massless ground states can be expanded as
\begin{equation}
\chi_i(q)  =  n_{i,s} (s) + n_{i,c} (c)   + \hbox{singlets} + \  \hbox{higher order in}\   q
\end{equation}
In most cases $(n_{i,s},n_{i,c})=(1,0)$ or $(0,1)$. In the diagonal MIPF these produce spinor-spinor
tensor products, leading to $N=2$ vector multiplets. 
For example, the tensor product $(3,3,3,3,3)$ has 4000 characters, of which 202 are relevant for the massless sector: the identity character, 100 characters with $n_s=1, n_c=0$,
their conjugates, with $n_s=0, n_c=1$ and one character with $n_s=n_c=1$, which is self-conjugate.
When diagonally
combined with itself, a character with $n_s\not=0$ and $n_c \not=0$ yields spinor-anti-spinor products, which
lead to $N=2$
hyper multiplets. For any MIPF, there is an additional hyper multiplet originating from the gravity sector, the square of the identity
characters. But characters with both spinors and anti-spinors are rare, and therefore in the diagonal MIPF the number of vector multiplets is usually much larger than the number of hyper multiplets. For example, for the
diagonal MIPF of the tensor product $(3,3,3,3,3)$ we get 101 vector multiplets and 1+1 hyper multiplet.  

However, the diagonal MIPF is not the one we use for building orientifolds. The simple current methods we use \cite{Fuchs:2000cm} to compute boundary and crosscap coefficients 
can only be used for symmetric MIPFs that are obtained by taking a simple current MIPF and multiplying it with the charge conjugation matrix $C$.  In the following we denote a generic MIPF as $Z$, and a simple current MIPF as $Z_S$.
Multiplication by $C$ does not affect the symmetry of a modular matrix ($(ZC)^T = C^T Z^T = CZ = ZC$ if $Z$ is symmetric, because $Z$ and $C$ commute).
There is no general orientifold formalism to deal with $Z$ itself, not even for the special case $Z=1$. 
This originates from the fact that 
Cardy's work \cite{Cardy:1989ir}, on which the entire formalism is based, dealt with the case
$Z=C$, where the Isihibashi states are in one-to-one correspondence with the RCFT primaries. Multiplying the torus partition functions $Z$ with $C$ has the
effect of conjugating the space-time spinors in one chiral sector. In the closed string sector this has the effect of interchanging the r\^ole of vector and
hyper-multiplets.  The MIPF defined by $C$ itself therefore yields usually a large number of $N=2$ hyper multiplets and few vector multiplets. For $(3,3,3,3,3)$, using the MIPF $C$, one gets
101+1 hyper multiplets and 1 vector multiplet. 

Compactification of a ten-dimensional type-II string on a Calabi-Yau manifold with Hodge numbers $(h_{11},h_{21})$ 
yields $h_{21}+1$ hyper- and $h_{11}$ vector multiplets for type IIA, and $h_{21}$ vector multiplets and $h_{11}+1$ hyper multiplets for type IIB. Hence if we
compare the spectra of the $(3,3,3,3,3)$ with compactifications on the quintic Calabi-Yau, with $(h_{11},h_{21}) = (1,101)$, then the diagonal MIPF matches 
a compactification of type-IIB on the quintic, while the charge conjugation MIPF matches compactification of type-IIA on the quintic. Since the charge conjugation
MIPF is the one to which we apply the orientifold projection, it is natural to use type-IIA language henceforth.
The Hodge numbers specified below are therefore obtained in the follow way.
From the number of $N=2$ vector multiplets we get the Hodge number $h_{11}$, while the number of $N=2$ hyper multiplets,
subtracting one universal one from the gravity sector, gives us the Hodge number $h_{21}$. We denote the corresponding Calabi-Yau manifold as $X$.

Note that in ten dimensions $Z$ and $ZC$ are identical (because the spinors are real), and since only symmetric $Z$ can be subject to an RCFT orientifold projection, this implies that only
type-IIB strings can be orientifold-projected. In four dimensions, $Z$ and $ZC$ are distinct, and since both are symmetric, both can be orientifold-projected. However, a general formula for boundary
and cross caps is only available for $Z_SC$. 

Geometrically, the partition function $Z$ can be interpreted either as type-IIB compactified on a manifold $X$ or type-IIA on the mirror of X, while $ZC$ corresponds
either to type-IIB on the mirror of $X$ or type-IIA on $X$. However, in the case of Gepner models the limitation to $ZC$ does not imply that we miss half of the possibilities. It turns out that  the set of
Gepner simple current MIPFs is closed under mirror symmetry, so that for any hodge pair $(h_{11},h_{12})=(p,q)$ there is a MIPF that yields $(h_{11},h_{12})=(q,p)$.

In the orientifold theory
the diagonal terms in the
 partition function  are subject to a Klein bottle projection. 
The full partition function has the form
\begin{equation}
\frac12   \left(  \sum_{ij}  Z_{ij} \chi_i (\tau) \chi_j(\bar \tau) + \sum_i K_i  \chi_i (2\tau)  \right)
\end{equation}
where $Z=Z_SC$.  

The Klein bottle projection works on the states with $i=j$. We can write these as
\begin{eqnarray*}
 \frac{1}{2}  \left[ Z_{ii} \chi_i (\tau) \chi_i (\tau) + K_i \chi_i (2\tau)\right]  = \ \ \ \ \ \ \ \ \ &  \\ 
 \left[Z_{ii}+K_i)\right]  \times  \left[ \frac{1}{2} ( \chi_i (\tau) \chi_i (\tau)  + \chi_i (2\tau)\right]&    \\  
+ \left[Z_{ii}-K_i)\right]  \times  \left[  \frac{1}{2} ( \chi_i (\tau) \chi_i (\tau)  - \chi_i (2\tau)\right]& 
\end{eqnarray*}
Now we can expand this in ground state spinors
\begin{eqnarray*}
\frac12 (\chi_i \chi_i \pm \chi_i) =& \\ (    n_{i,s} n_{i,s} \pm n_{i,s} ) (s \otimes s)_{\rm S} &  + \ \  ( n_{i,s} n_{i,s} \mp n_{i,s} ) (s \otimes s)_{\rm A}\\
                                             +\  ( n_{i,c} n_{i,c} \pm n_{i,c} ) (c \otimes c)_{\rm S} &  + \ \ ( n_{i,c} n_{i,c} \mp n_{i,c} ) (c \otimes c)_{\rm A}\\
                                              +\ n_{i,c} n_{i,s}\  s \otimes c&
\end{eqnarray*}
Here ``S" means Symmetric and ``A" anti-symmetric, and this refers to the sign of the projection obtained from the Klein
bottle signs. For fermions, there is an additional spin statistics sign flipping the projection. Therefore the symmetric
projection ($K_i=+1)$ of an $N=2$ vector superfield yields an $N=1$ chiral multiplet, and the anti-symmetric projection
($K_i=-1$) yields and $N=2$ vector field, even though group-theoretically the vector is contained in the {\it symmetric} part of
the lightcone spinor-spinor tensor product. We denote the corresponding components of $h_{11}$ as $h_{11}^+$  and $h_{11}^-$, where
the subscript denote the sign of $K_i$. Hence by $h_{11}^-$ we mean the anti-symmetric Klein bottle projection, yielding the
symmetric light-cone spinor-spinor product, which produces a vector boson. Then $h_{11}^-$ is identified with the number of
vector bosons in the closed string spectrum, and $h_{11}^+$ with the number of chiral multiplets. Here we are following the
conventions chosen in \cite{Dijkstra:2004cc}. In the literature these superscripts are sometimes flipped, so it is better to
directly compare the number of vector bosons.
Note that $h_{21}$ will get contributions from  $(Z_{ii}+K_i)$ and from $(Z_{ii}-K_i)$, so one could split $h_{21}$ into plus and
minus contributions, if one wishes to do so. However, both projections of $h_{21}$ yield chiral multiplets, and so we do
not make the distinction.
In addition to the projected diagonal contributions there are off-diagonal terms which may contribute to both $h_{21}$ and $h_{11}$. 
The latter can be split into equal numbers of symmetric and anti-symmetric contributions ($N=1$ chiral and vector multiplets).
The vector multiplets arising from diagonal and off-diagonal terms are added, and  this then defines the quantity $h_{11}^-$.
This is subtracted from $h_{11}$ to obtain $h_{11}^+$. 

Applying this to the example of the quintic, we get the following. The formalism of \cite{Fuchs:2000cm} allows only one
Klein bottle projection in this case, which is always symmetric. Therefore the single $N=2$ vector boson of the type-IIA theory  
yields a single chiral multiplet and no vector multiplets. The 101+1 $N=2$ hyper multiplets yield 101+1 chiral multiplets. Each $N=2$ hyper multiplet contains two $N=1$ chiral multiplets, but there is also an overall factor $\frac12$ in the projection.

 The states that can propagate in the transverse channel of the annulus, Moebius strip or Klein bottle 
 (``Ishibashi states") originate from the
 terms $Z_{ii^c} \not=0$ in the partition function.  If $i$ represents a massless character, the ground state contains a 
 massless spinor, and then $Z_{ii^c}$ contributes to $h_{12}$. Massless Ishibashi states give rise to axions and 
and also to tadpole conditions. 
 Hence the smaller $h_{21}$ is, the smaller the number of tadpole
 conditions, and the smaller the number of axions (note however that there are in general also contributions to $h_{12}$ from
terms $Z_{ij}, j\not=i^c$, as well as non-trivial multiplicities $n_{i,s} > 1, n_{i,c} > 1$, which do not give rise to additional
axions or tadpole equations). 
 The number of axions determines first of all the likelihood of solving the constraint that $Y$ remains massless.
 Indeed, in \cite{Anastasopoulos:2006da} it was observed that standard model configurations are most frequently found for compactifications with small $h_{21}$, whereas in 
\cite{Dijkstra:2004ym,Dijkstra:2004cc} it was observed that most tadpole solution are found for small $h_{21}$\rlap.\footnote{Note that different conventions were used in these papers
for the Hodge numbers. In \cite{Dijkstra:2004ym} the same convention is used as in the present paper, namely quoting the Hodge numbers of the compactification manifold 
corresponding to the MIPF $Z$, whereas in \cite{Anastasopoulos:2006da} and \cite{Dijkstra:2004ym} the Hodge numbers derived from the torus  $ZC$ were quoted. This
amounts to an interchange of $h_{11}$ and $h_{12}$.}

 \newcommand\wb{\,\linebreak[0]} \def\wB {$\,$\wb}
 \newcommand\Bi[1]    {\bibitem{#1}}
 \newcommand\J[5]   {{\sl #5}, {#1} {#2} ({#3}) {#4} }
 \newcommand\Prep[2]  {{\sl #2}, preprint {#1}}
 \def\jf    {J.\ Fuchs}
 \def\adma  {Adv.\wb Math.}
 \def\anop  {Ann.\wb Phys.}
 \def\aspm  {Adv.\wb Stu\-dies\wB in\wB Pure\wB Math.}
 \def\atmp  {Adv.\wb Theor.\wb Math.\wb Phys.}
 \def\comp  {Com\-mun.\wb Math.\wb Phys.}
 \def\ijmp  {Int.\wb J.\wb Mod.\wb Phys.\ A}
 \def\jhep  {J.\wb High\wB Energy\wB Phys.}
 \def\mpla  {Mod.\wb Phys.\wb Lett.\ A}
 \def\nuci  {Nuovo\wB Cim.}
 \def\nupb  {Nucl.\wb Phys.\ B}
 \def\phlb  {Phys.\wb Lett.\ B}
 \def\phrl  {Phys.\wb Rev.\wb Lett.}
 \def\NH     {{North Holland Publishing Company}}
 \def\SV     {{Sprin\-ger Ver\-lag}}
 \def\WS     {{World Scientific}}
 \def\Ad     {{Amsterdam}}
 \def\Be     {{Berlin}}
 \def\Si     {{Singapore}}

\begin{landscape}

\begin{table}[p] \centering\footnotesize
\begin{tabular}{|c|c|c|c|c|c|c|c|c|c|c|c|c|c|c|c|c|c|c|c|c|c|c|c|c|c|}
\hline
    Nr        & U/S   &  $U(1)$  & {\bf ab}    & {\bf ab}$^*$ & {\bf a}$^*${\bf c} & {\bf a}$^*${\bf c}$^*$ & {\bf a}$^*${\bf d} & {\bf a}$^*${\bf d}$^*$   & {\bf bd}$^*$      & {\bf b}$^*${\bf d}$^*$ & {\bf c}$^*${\bf d}      &  {\bf c}{\bf d} & {\bf bc}              & {\bf bc}$^*$    & Total   &  $\mathbb{Z}_2$  & $\mathbb{Z}_3$  & Tadp.    \\
 \hline
               &          &                 & Q           &Q                     & U$^c$                  & D$^c$                         &  D$^c$                       &  U$^c$                          &       L                     &  L                                 & E$^c$                       & N$^c$           & H$_d$/L               & H$_u$  & & & & \\
 \hline
 7506   & S      & 1               & 3            & $-$                 & 3                            & 3                                  & 0                                  & 0                               &  3                               &     $-$                          &       3                            &         3            &        0                     &         0         & 40590   &    2152    &            16     & 320  ($\mathbb{Z}_2$)  \\
 2751   & S      & 2              & 3            & $-$                 & 3                             & 3                                  & 0                                  & 0                               &  3                              &     $-$                           &       3                             &         3            &        0                    &         0       & 869428   &    0    &          59808     & 41136      \\
 \hline
 14704 & S      & 1              & 3            & $-$                 & 1                             & 2                                  & 1                                  & 2                               &   0                              &   $-$                            &       3                            &        0            &        3                     &         0           & 380 &     0         &               0    & 0 \\
 14062 & S      &  1               &  3            & $-$                 &    2                          &    2                             &     1                            & 1                                &    2                         &   $-$                                  &      3                            &            1        &        1                         &       0   &   304           &     0        &     0              & 0   \\
  8745   & S      &   1             &   3            & $-$                 &    3                     &      2                              &     1                               &            0                      &   4                          &   $-$                               &         3                           &         2            &             0                     &     1      &        92        &       0       &         0       & 0      \\
 11196  & S      &    1           &   3           & $-$                 &    3                         &   4                                    &     -1                         &        0                             &    2                         &   $-$                               &     3                               &    4                  &       1                    &      0        &     40         &      0       &         0          & 0       \\
 \hline
 10551   & U      &    1           &     1        &    2                &      3                   &       3                                  &              0                          &              0                         &        3             &          0                                   &        3                        &          3              &    0                     &        0        & 116   &       0       &          0   & 0  \\
 1352        & U      &    2           &  1          & 2                   &   3                       &    3                                   &         0                             &         0                             &      3                        &     0                                        &   3                               &      3                    &     0                       &     0      &    20176    & 0        &       1472        & 0     \\
  13058    & U      &   1           &     1        &        2             &      3                   &       3                                 &       0                                &        0                              &    1                     &       2                                   &      3                     &      3               &        2                 &    2     &    68    & 0        &       0       & 0   \\
  7573       & U      &     2            &  1           &  2                &   3                      &     3                               &      0                                  &    0                               &    1                     &     2                                    &        3                        &   3                     &      2                &        2    &     14744      &   0           &      0    & 0       \\
 16074   & U      &     1          &      0         &       3              &     3                        &      3                                   &       0                                 &      0                                 &      3                  &        0                                   &          3                      &         3              &       3                   &      3      &     128       &   0           &      0     & 0   \\
 7967      & U      &       2          &     0         &        3             &    3                        &   3                             &             0                          &                  0                    &           3             &   0                                   &     3                           &    3                   &     3                      &   3           &       5856          &    0         &      0          & 0  \\
   12106     & U      &   1            &    1          &   2                  &     3                        &     3                                    &      0                                  &      0                                 &      2               &      1                                       &      3                          &     3                   &     1                   &    1       &      32          &     0         &        0     & 0        \\
  7976      & U      &      2          &      1        &       2             &    3                        &      3                                 &       0                                 &          0                             &   2                    &    1                                        &   3                            &    3                  &  1                       & 1        &          5764       &     0        &       192     & 0         \\
   13844    & U      &     2           &     1          &    2                 &      3                      &       3                                 &          0                              &        0                               &   0                     &            3                                &   3                             &   3                 &      3                   &     3        &     1096         &    0          &           0       & 0  \\
 14793    & U      &   2           &      2      &         1           &            3                &        3                            &                0                        &                  0                     &        4             &      -1                                      &            3                   &           3            &    -1                     &     -1      &        400         &     0        &          0    & 0   \\
 13762    & U      &    2            &   0         &        3             &      3                    &      3                              &               0                        &                  0                    &     6               &       -3                                     &       3                      &       3              &           0               &        0        &       320         &      0        &         0   & 0    \\
  14850    & U      &    2            &   0           &      3              &     3                      &    3                                  &           0                            &            0                          &      4                &     -1                                      &      3                        &      3               &       2                  &    2      &     96            &      0       &        0      & 0   \\
   14792    & U      &    2          &     0          &      3                &    3                       &     3                                 &     0                                   &         0                            &        0                 &             3                              &      3                         &     3               &      6            &       6      &     32            &    0         &       32  & 0  \\
\hline
   7488     & U      &     1          &     1        &         2             &        1                    &   2                                      &     1                                   &  2                                    &    0                    &             0                               &   3                          &     0                  &      3                  &   0       &       2864         & 0 &    144          & 0   \\
13015    & U      &   2            &     1         &      2               &     1                    &  2                                  &    1                                  &     2                                 &        0                &                 0                           &     3                           &  0                     &         3                 &     0        &   352      &      0       &     0     & 0     \\
  18086    & U      &    1           &     2         &     1              &      2                    &      4                                &     -1                                 &      1                               &          0               &            0                                 &      3                      &        3              &        3                   &    0       &     68           &       0       &        0   & 0     \\
   13644        & U      &     1          &       0        &      3                &     1                        &        3                                 &   0                           &      2                                           &    1            &         -2                                    &   3                             &       1                 &          5               &   1        &      8          &      0        &       0         & 0  \\
   653      & U      &         1        &        0       &         3             &           0                  &                3                         &           0                             &         3                              &     0                    &           -3                                 &             3                   &     0                &     6                  &      0           &          4       &     0        &           0      & 0    \\
\hline \hline    
 
 \hline
\end{tabular}
\caption{\small {Chiral Spectra of the 24 classes of models. Column 1 specifies the number assigned to the
spectrum class in the database, column 2 indicates if the {\bf b}-brane is unitary or symplectic,
 column 3 lists the number of massless $U(1)$'s, including $Y$, and the subsequent
columns list the chiral brane intersections. The last three columns indicate the total number of spectra in each class that is present in the database, and
the total number with a certain discrete symmetry. The last column indicates in how many cases there was also a solution to the tadpole conditions.}} 
\label{Madrid}
\end{table}\vspace{5pt}

\begin{table}[p] \centering
\begin{tabular}{|l|l|c|c|c|c|c|c|c|c|c|c|c|c|c|c|c|c|c|c|c|c|c|c|c|c|}
\hline
    Nr        & Type   &  $U(1)$  & $A_{\bf a}$    & ${\bf a}^*{\bf b}$ & ${\bf a}^*{\bf b}^*$ & ${\bf ac}$ & ${\bf bc}$ &  ${\bf bc}^*$ & $A_2$ & $S_2$& $A_3$ & $S_3$  & Total & $\mathbb{Z}_2$ &   $\mathbb{Z}_3$  & Tadp.\\
    \hline
    7        & UO      &    1            &     3                &    3                            &         $-$       &       $-$     &    $-$          &    $-$                          &   $-$        &  $-$           &     $-$        &     $-$            &   16845      &       0              &        0                     &  0  \\
    218     & UU      &    2            &     3              &    3                            &         0           &       $-$     &    $-$          &    $-$                          &   0            &   -3           &     $-$        &     $-$            &   1049      &        0             &            0                 &  0  \\
    345     & UU      &    1            &     3              &    3                            &         0           &       $-$     &    $-$          &    $-$                          &   0            &   -3           &     $-$        &     $-$            &   1136      &      18               &        0                     &  0  \\
    742     & UU      &    1            &     3              &    2                            &         1           &       $-$     &    $-$          &    $-$                           &   0            &   -1           &     $-$        &     $-$            &   146      &       0              &          0                   &  0  \\
    18371     & UU      &    1        &     3              &    6                           &        -3           &       $-$     &    $-$          &    $-$                          &   0            &   -9           &     $-$        &     $-$            &   12      &         0            &    0                         &    0 \\ \hline
    57     & UUO      &    1            &     3              &    3                           &         3           &       3         &   0            &    0                            &   0            &   0           &     $-$        &     $-$            &   13402      &       552             &       0                      &  0 \\
    998     & UUO      &    2            &     3              &    3                           &         3           &       3         &   0            &    0                            &   0            &   0           &     $-$        &     $-$            &   18890      &      0               &         0                    &  0  \\
    1000     & UUU      &    3           &     3              &    3                           &         3           &       3         &   -3            &    3                            &   0            &   0       &        3     &    0            &   7276      &           0          &        0                     &  0 \\
    4004     & UUU      &    2           &     3              &    3                           &         3           &       3         &   -3            &    3                            &   0            &   0       &        3     &    0            &   1706      &           4         &          0                 &   0 \\
    4316     & UUU      &    2           &     3              &    3                           &         3           &       3         &   -3            &    3                            &   0            &   0       &        3     &    0            &   5236      &        180          &       120                      &  0 \\
    4324     & UUU      &    1           &     3              &    3                           &         3           &       3         &   -3            &    3                            &   0            &   0       &        3     &    0            &   1278      &          8        &           0                  &  0 \\
    4325     & UUU      &    1           &     3              &      3                         &           3           &          3        &       -4     &        4                            &       0         &   0          &        4        &      1           &   96      &     48              &          0                   &  0\\
  \hline   
\end{tabular}
\caption{\small {Chiral Spectra of the $SU(5)$ GUT models considered here. Here {\bf a} denotes the $U(5)$ brane,
and {\bf b} and {\bf c} the additional branes. Column 2 specifies the brane types, and column 3 the number of massless $U(1)$'s including $Y$ (which is embedded entirely in $SU(5)$. The 
difference between 4004 and 4316 is the embedding of the additional $U(1)$, which is $Y_{\bf b}$ in  nr. 4004 and $Y_{\bf a}-5Y_{\bf c}$  in nr. 4316.}} 
\label{SU5}
\end{table}

\end{landscape}

\end{document}